\documentclass[aps,prd,superscriptaddress,showpacs]{revtex4}

\usepackage{graphicx}
\usepackage{dcolumn}
\usepackage{bm}
\usepackage{multirow}

\newcommand{\inmath}[1] {\ifmmode#1\else$#1$\fi}
\newcommand{\definmath}[2] {\def#1{\ifmmode#2\else$#2$\fi}}
\newcommand{\ppbar}{p\bar{p}}

\newcommand{\ttbar}{t\bar{t}}
\newcommand{\bbbar}{b\bar{b}}
\newcommand{\ccbar}{c\bar{c}}
\newcommand{\tptm}{\mbox{$\tau^+\tau^-$}}
\newcommand{\met}{E\!\!\!/_T }

\newcommand{\GeV}{\mbox{$\mathrm{GeV}$}}
\newcommand{\TeV}{\mbox{$\mathrm{TeV}$}}
\newcommand{\GeVc}{\mbox{$\mathrm{GeV}\!/\!c$}}
\newcommand{\GeVcc}{\mbox{$\mathrm{GeV}\!/\!c^2$}}
\newcommand {\ra}         {\rightarrow}
\definmath{\Pt}      {P_{\mathrm T}}
\definmath{\Et}      {E_{\mathrm T}}
\definmath{\Ht}      {H_{\mathrm T}}
\newfont{\sans}{cmr8 at 10pt}
\begin{document}

%

\hfill FERMILAB-PUB-05-220-E\\


\title{Measurement of the {\boldmath $\ttbar$} Production Cross Section in {\boldmath $\ppbar$} Collisions
at {\boldmath $\sqrt{s}=1.96~\TeV$} Using Lepton Plus Jets Events
with Semileptonic B Decays to Muons}

\vspace*{1.0cm}

\affiliation {Institute of Physics, Academia Sinica, Taipei, Taiwan
11529,Republic of China} \affiliation {Argonne National Laboratory,
Argonne, Illinois 60439} \affiliation {Institut de Fisica d'Altes
Energies, Universitat Autonoma de Barcelona, E-08193, Bellaterra
(Barcelona), Spain} \affiliation {Istituto Nazionale di Fisica
Nucleare, University of Bologna, I-40127 Bologna, Italy}
\affiliation {Brandeis University, Waltham, Massachusetts 02254}
\affiliation {University of California, Davis, Davis, California
95616} \affiliation {University of California, Los Angeles, Los
Angeles, California  90024} \affiliation {University of California,
San Diego, La Jolla, California  92093} \affiliation {University of
California, Santa Barbara, Santa Barbara, California 93106}
\affiliation{Instituto de Fisica de Cantabria, CSIC-University of
Cantabria, 39005 Santander, Spain} \affiliation{Carnegie Mellon
University, Pittsburgh, PA  15213} \affiliation{Enrico Fermi
Institute, University of Chicago, Chicago, Illinois 60637}
\affiliation{Joint Institute for Nuclear Research, RU-141980 Dubna,
Russia} \affiliation{Duke University, Durham, North Carolina  27708}
\affiliation{Fermi National Accelerator Laboratory, Batavia,
Illinois 60510} \affiliation{University of Florida, Gainesville,
Florida  32611} \affiliation{Laboratori Nazionali di Frascati,
Istituto Nazionale di Fisica Nucleare, I-00044 Frascati, Italy}
\affiliation{University of Geneva, CH-1211 Geneva 4, Switzerland}
\affiliation{Glasgow University, Glasgow G12 8QQ, United Kingdom}
\affiliation{Harvard University, Cambridge, Massachusetts 02138}
\affiliation{Division of High Energy Physics, Department of Physics,
University of Helsinki and Helsinki Institute of Physics, FIN-00014,
Helsinki, Finland} \affiliation{Hiroshima University,
Higashi-Hiroshima 724, Japan} \affiliation{University of Illinois,
Urbana, Illinois 61801} \affiliation{The Johns Hopkins University,
Baltimore, Maryland 21218} \affiliation{Institut f\"{u}r
Experimentelle Kernphysik, Universit\"{a}t Karlsruhe, 76128
Karlsruhe, Germany} \affiliation{High Energy Accelerator Research
Organization (KEK), Tsukuba, Ibaraki 305, Japan} \affiliation{Center
for High Energy Physics: Kyungpook National University, Taegu
702-701; Seoul National University, Seoul 151-742; and SungKyunKwan
University, Suwon 440-746; Korea} \affiliation{Ernest Orlando
Lawrence Berkeley National Laboratory, Berkeley, California 94720}
\affiliation{University of Liverpool, Liverpool L69 7ZE, United
Kingdom} \affiliation{University College London, London WC1E 6BT,
United Kingdom} \affiliation{Massachusetts Institute of Technology,
Cambridge, Massachusetts  02139} \affiliation{Institute of Particle
Physics: McGill University, Montr\'{e}al, Canada H3A~2T8; and
University of Toronto, Toronto, Canada M5S~1A7}
\affiliation{University of Michigan, Ann Arbor, Michigan 48109}
\affiliation{Michigan State University, East Lansing, Michigan
48824} \affiliation{Institution for Theoretical and Experimental
Physics, ITEP, Moscow 117259, Russia} \affiliation{University of New
Mexico, Albuquerque, New Mexico 87131} \affiliation{Northwestern
University, Evanston, Illinois  60208} \affiliation{The Ohio State
University, Columbus, Ohio  43210} \affiliation{Okayama University,
Okayama 700-8530, Japan} \affiliation{Osaka City University, Osaka
588, Japan} \affiliation{University of Oxford, Oxford OX1 3RH,
United Kingdom} \affiliation{University of Padova, Istituto
Nazionale di Fisica Nucleare, Sezione di Padova-Trento, I-35131
Padova, Italy} \affiliation{University of Pennsylvania,
Philadelphia, Pennsylvania 19104} \affiliation{Istituto Nazionale di
Fisica Nucleare Pisa, Universities of Pisa, Siena and Scuola Normale
Superiore, I-56127 Pisa, Italy} \affiliation{University of
Pittsburgh, Pittsburgh, Pennsylvania 15260} \affiliation{Purdue
University, West Lafayette, Indiana 47907} \affiliation{University
of Rochester, Rochester, New York 14627} \affiliation{The
Rockefeller University, New York, New York 10021}
\affiliation{Istituto Nazionale di Fisica Nucleare, Sezione di Roma
1, University di Roma ``La Sapienza," I-00185 Roma, Italy}
\affiliation{Rutgers University, Piscataway, New Jersey 08855}
\affiliation{Texas A\&M University, College Station, Texas 77843}
\affiliation{Texas Tech University, Lubbock, Texas 79409}
\affiliation{Istituto Nazionale di Fisica Nucleare, University of
Trieste/\ Udine, Italy} \affiliation{University of Tsukuba, Tsukuba,
Ibaraki 305, Japan} \affiliation{Tufts University, Medford,
Massachusetts 02155} \affiliation{Waseda University, Tokyo 169,
Japan} \affiliation{Wayne State University, Detroit, Michigan
48201} \affiliation{University of Wisconsin, Madison, Wisconsin
53706} \affiliation{Yale University, New Haven, Connecticut 06520}
\author{D.~Acosta}
\affiliation{University of Florida, Gainesville, Florida  32611}
\author{J.~Adelman}
\affiliation{Enrico Fermi Institute, University of Chicago, Chicago,
Illinois 60637}
\author{T.~Affolder}
\affiliation {University of California, Santa Barbara, Santa
Barbara, California 93106}
\author{T.~Akimoto}
\affiliation{University of Tsukuba, Tsukuba, Ibaraki 305, Japan}
\author{M.G.~Albrow}
\affiliation{Fermi National Accelerator Laboratory, Batavia,
Illinois 60510}
\author{D.~Ambrose}
\affiliation{Fermi National Accelerator Laboratory, Batavia,
Illinois 60510}
\author{S.~Amerio}
\affiliation{University of Padova, Istituto Nazionale di Fisica
Nucleare, Sezione di Padova-Trento, I-35131 Padova, Italy}
\author{D.~Amidei}
\affiliation{University of Michigan, Ann Arbor, Michigan 48109}
\author{A.~Anastassov}
\affiliation{Rutgers University, Piscataway, New Jersey 08855}
\author{K.~Anikeev}
\affiliation{Fermi National Accelerator Laboratory, Batavia,
Illinois 60510}
% ...

\collaboration{CDF Collaboration}
\noaffiliation

\date{\today}

\begin{abstract}

We present a measurement of the $\ttbar$ production cross section
using $194\,\mathrm{pb^{-1}}$ of CDF~II data using events with a
high transverse momentum electron or muon, three or more jets, and
missing transverse energy. The measurement assumes 100\%
$t\rightarrow Wb$ branching fraction.  Events consistent with
$\ttbar$ decay are found by identifying jets containing heavy flavor
semileptonic decays to muons. The dominant backgrounds are evaluated
directly from the data. Based on 20 candidate events and an expected
background of 9.5$\pm$1.1 events, we measure a production cross
section of $5.3\pm3.3^{+1.3}_{-1.0}\,\mathrm{pb}$, in agreement with
the standard model.

\vspace{0.5cm}
\end{abstract}


\maketitle
\section{\label{sec:Intro}Introduction}
Top quark pair production in the standard model proceeds via either
quark-antiquark annihilation or gluon-gluon fusion.  At the Fermilab
Tevatron collider, with a center-of-mass energy of $1.96~\TeV$,
quark-antiquark annihilation is expected to dominate.  For a top
mass of $175~\GeVcc$, the calculated cross section is $6.7^{+0.7}_{-0.9}~$pb~\cite{theory},
and is approximately 0.2~pb smaller for each $1~\GeVcc$ increase in the value
of the top mass over the range $170~\GeVcc<M_{\mathrm{top}}<190~\GeVcc$.

Measurements of the cross section for top quark pair production
provide a test of  QCD, as well as the standard model decay
$t\rightarrow Wb$. Non-standard model production mechanisms, such as
the production and decay of a heavy resonance into $\ttbar$
pairs~\cite{Xtt}, could enhance the measured cross section.
Non-standard model top quark decays, such as the decay into
supersymmetric particles~\cite{susy}, could suppress the measured
value, for which a $t\rightarrow Wb$ branching fraction of nearly
100\% is assumed.

In this paper we report a measurement of the $\ttbar$ production
cross section in $\ppbar$ collisions at $\sqrt{s}=1.96~\TeV$ with
the CDF~II detector at the Fermilab Tevatron.  The standard model
decay $t\rightarrow Wb$ of the top quark results in a final state
from $\ttbar$ production of two $W$ bosons and two bottom quarks. We
select events consistent with a decay of one of the $W$ bosons to an
electron or muon plus a neutrino, both of which have large momentum
transverse to the beam direction ($\Pt$).  We refer to these high
$\Pt$ electrons or muons as the ``primary lepton". The neutrino is
undetected and results in an imbalance in transverse momentum. The
imbalance is labeled ``missing $\Et$" ($\met$) since it is
reconstructed based on the flow of energy in the
calorimeter~\cite{MET}.  The other $W$ boson in the event decays to
a pair of quarks.  The two quarks from the $W$ boson and the two $b$
quarks from the top decays hadronize and are observed as jets of
charged and neutral particles. This mode is referred as $W$ plus
jets. We take advantage of the semileptonic decay of $b$ hadrons to
muons to identify final-state jets that result from hadronization of
the bottom quarks. Such a technique, called ``soft-lepton tagging"
(SLT), is effective in reducing the background to the $\ttbar$
signal from $W$ boson produced in association with several hadronic
jets with large transverse momentum. The production cross section is
measured in events with three or more jets and at least one SLT
tagged jet.

This measurement is complementary to other measurements from CDF~II,
which use secondary vertex tagging, kinematic fitting, or a combination
of the two~\cite{SVX}~\cite{kinPRD}~\cite{SVXkin}.  A forthcoming
paper~\cite{combXsec} will present a combined cross section measurement based on
the result of these four analyses.

Previous measurements~\cite{RunI} from Run~I at the Tevatron have
measured production cross sections statistically consistent with the
standard model prediction.  This and other Run~II measurements are
made at a slightly higher center of mass energy ($1.96~\TeV$ vs.\@
$1.8~\TeV$) and with nearly twice as much integrated luminosity.

The organization of this paper is as follows:
Section~\ref{sec:CDFdet} reviews the detector systems relevant to
this analysis. The trigger and event selection, the data and the
Monte Carlo samples and the SLT tagging algorithm are described in
Section~\ref{sec:Data}. The estimate of the background is presented
in Section~\ref{sec:Bak}. The acceptance and the $\ttbar$ event
tagging efficiency are described in Section~\ref{sec:Acc}. The
evaluation of the systematic uncertainties on the measurement is
presented in Section~\ref{sec:sys}. The $\ttbar$ production cross
section measurement and the conclusions are presented in
Section~\ref{sec:results} and  Section~\ref{sec:concl}.

\section{\label{sec:CDFdet}The CDF~II Detector}
\begin{figure}[htbp]
\begin{center}
\includegraphics[width=9in]{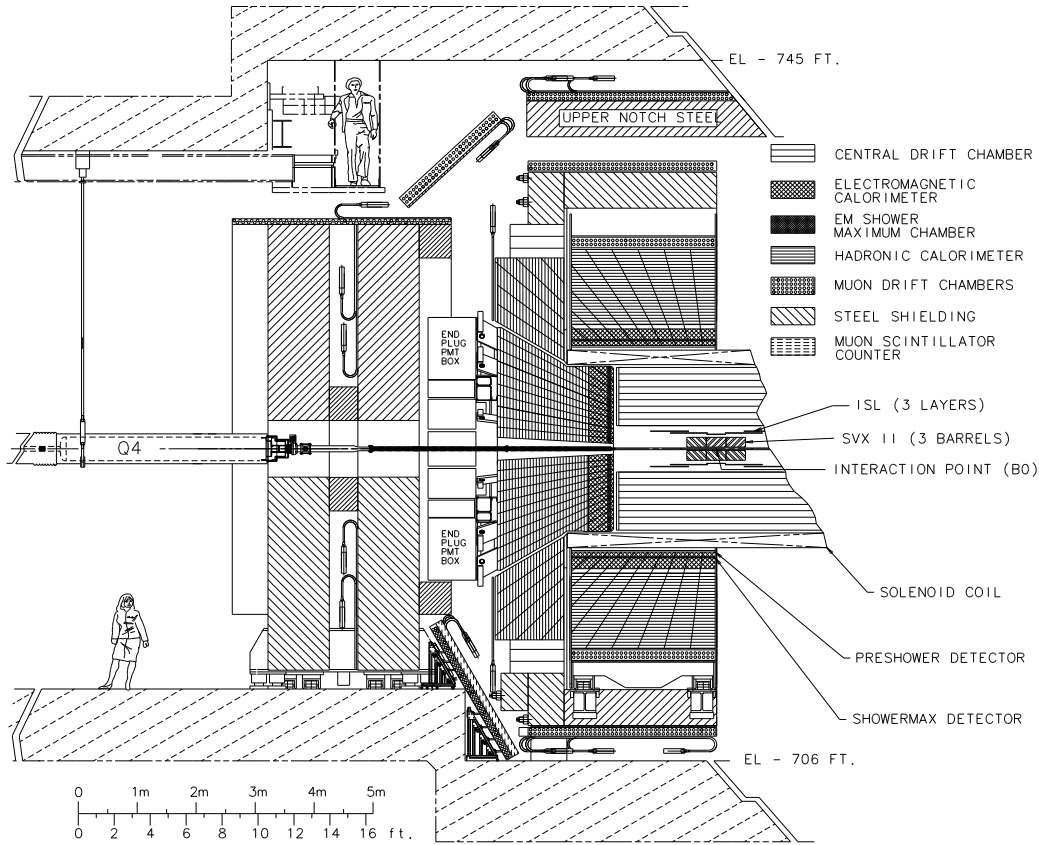}
\caption{Elevation view of the CDF~II detector.} \label{fig:cdfel}
\end{center}
\end{figure}

The CDF~II detector is described in detail in~\cite{CDF}, only the
components relevant to this measurement are summarized here.
The CDF~II detector is a nearly azimuthally and forward-backward
symmetric detector designed to study $\ppbar$ interactions at the
Fermilab Tevatron.  It consists of a magnetic spectrometer
surrounded by calorimeters and muon chambers.  An elevation view of
the CDF~II detector is shown in Figure~\ref{fig:cdfel}.

Charged particles are tracked inside a 1.4~T solenoidal magnetic
field by an 8-layer silicon strip detector covering radii from
1.5~cm to 28~cm, followed by the central outer tracker (COT), an
open-cell drift chamber that provides up to 96 measurements of
charged particle position over the radial region from 40~cm to
137~cm. The 96 COT measurements are arranged in 8 ``superlayers" of
12 sense wires each alternating between axial and 2$^\circ$ stereo.
The COT and the silicon detectors track charged particles for
$|\eta|<1$ and $|\eta|<2$, respectively.

Surrounding the tracking system are electromagnetic and hadronic
calorimeters, used to measure charged and neutral particle energies.
The electromagnetic calorimeter is a lead-scintillator sandwich and
the hadronic calorimeter is an iron-scintillator sandwich.  Both
calorimeters are segmented in azimuth and polar angle to provide
directional information for the energy deposition.  The segmentation
varies with position on the detector and is 15$^\circ$ in azimuth by
0.1 units of $\eta$ in the central region ($|\eta|<1.1$).
Segmentation in the plug region ($1.1<|\eta|<3.6$) is 7.5$^\circ$ up to
$|\eta|<2.1$, and 15$^\circ$ for $|\eta|>2.1$ in azimuth, while
ranging from 0.1 to 0.64 units of $\eta$ in pseudo-rapidity (a
nearly constant 2.7$^\circ$ change in polar angle).  The
electromagnetic calorimeters are instrumented with proportional and
scintillating strip detectors that measure the transverse profile of
electromagnetic showers at a depth corresponding to the shower
maximum.

Outside the central calorimeter are four layers of muon drift
chambers covering $|\eta|<0.6$ (CMU).  The calorimeter provides
approximately 1 meter of steel shielding. Behind an additional 60~cm
of steel in the central region sit an additional four layers of muon
drift chambers (CMP) arranged in a box-shaped layout around the
central detector. Central muon extension (CMX) chambers, which are
arrayed in a conical geometry, provide muon detection for the region
$0.6<|\eta|<1$ with four to eight layers of drift chambers,
depending on polar angle. All the muon chambers measure the
coordinates of hits in the drift direction, $x$, via a drift time
measurement and a calibrated drift velocity. The CMU and the CMX
also measure the longitudinal coordinate, $z$. The longitudinal
coordinate is measured in the CMU by comparing the height of pulses,
encoded in time-over-threshold, at opposite ends of the sense wire.
In the CMX, the conical geometry provides a small stereo angle from
which the $z$ coordinate of track segments can be determined.
Reconstructed track segments have a minimum of three hits, and a
maximum of four hits in the CMU and the CMP, and 8 hits in the CMX.

\section{\label{sec:Data}Data Sample and Event Selection}
In this section we describe the collision data and the Monte Carlo
samples used in this analysis. Section~\ref{sec:CollBeam} outlines
the trigger system used for the initial selection of events from the
$\ppbar$ collisions. Section~\ref{sec:MCdata} describes the Monte
Carlo samples used for acceptance and background studies. The
selection of the $W$+jets datasets from the triggered data samples
is presented in Section~\ref{sec:Wplusjets}. The $\ttbar$ signal is
extracted from the $W$+jets events through the identification of
candidate $b$ hadron semileptonic decays to muons. The algorithm for
identifying these decays is summarized in Section~\ref{sec:SLT}, and
its application to the $W$+jets dataset is described in
Section~\ref{sec:Evsel}.

 This analysis is based on an integrated luminosity of
$194\pm11$~pb$^{-1}$~\cite{klimenko} collected with the CDF~II
detector between March 2002 and August 2003 (175~pb$^{-1}$ with the
CMX detector operational).

\subsection{\label{sec:CollBeam}{\boldmath$ p\bar p$} Collision Data}
CDF~II employs a three-level trigger system, the first two
consisting of special purpose hardware and the third consisting of a
farm of commodity computers.  Triggers for this analysis are based
on selecting high transverse momentum electrons and muons.  The
electron sample is triggered as follows:  At the first trigger level
events are selected by requiring a track with $\Pt>8~\GeVc$ matched
to an electromagnetic calorimeter tower with $\Et>8~\GeV$ and little
energy in the hadronic calorimeter behind it.  At the second trigger
level, calorimeter energy clusters are assembled and the track found
at the first level is matched to an electromagnetic cluster with
$\Et>16~\GeV$.  At the third level, offline reconstruction is
performed and an electron candidate with $\Et>18~\GeV$ is required.
The efficiency of the electron trigger is measured from
$Z\rightarrow ee$ data and found to be $(96.2\pm
0.6)\%$~\cite{WZPRD}. The selection of the muon sample begins at the
first trigger level with a track with $\Pt>4~\GeVc$ matched to hits
in the CMU and the CMP chambers or a track with $\Pt>8~\GeVc$
matched to hits in the CMX chambers. At the second level, a track
with $\Pt>8~\GeVc$ is required in the event for about 70\% of the
integrated luminosity, while for the remainder, triggers at the
first level are fed directly to the third level.  At the third
trigger level, a reconstructed track with $\Pt>18~\GeVc$ is required
to be matched to the muon chamber hits. The efficiency of the muon
trigger, measured from $Z\rightarrow \mu\mu$ data, is $(88.7\pm
0.7)\%$ for CMU/CMP muons and $(95.4\pm 0.4)\%$ for CMX
muons~\cite{WZPRD}.

\subsection{\label{sec:MCdata}Monte Carlo Datasets}
The detector acceptance of $\ttbar$ events is modeled using {\tt PYTHIA}
v6.2~\cite{Pythia} and {\tt HERWIG} v6.4~\cite{Herwig}.  These are
leading-order event generators with parton showering to simulate radiation
and fragmentation effects.  The generators are used with the CTEQ5L
parton distribution functions~\cite{CTEQ5L}. Decays of $b$ and $c$
hadrons are modeled using {\tt QQ} v9.1~\cite{QQ}.   Estimates of
backgrounds from diboson production ($WW$, $WZ$, $ZZ$) are derived using the
{\tt ALPGEN} generator~\cite{Alpgen} with parton showering provided
by {\tt HERWIG}. The background from single top production (eg.
$W^*\rightarrow t\overline{b}$) is simulated using {\tt PYTHIA}. Samples of the
remaining backgrounds are derived directly from the data as
described in Section~\ref{sec:Bak}.

The detector simulation reproduces the response of the detector to
particles produced in $\ppbar$ collisions. The same detector
geometry database is used in both the simulation and the
reconstruction, and tracking of particles through matter is
performed with {\tt GEANT3}~\cite{geant}.  The drift model for the
COT uses a parametrization of a {\tt GARFIELD}
simulation~\cite{garfield} with parameters tuned to match COT
collider data. The calorimeter simulation uses the {\tt
GFLASH}~\cite{gflash} parametrization package interfaced with {\tt
GEANT3}.  The {\tt GFLASH} parameters are tuned to test beam data
for electrons and high-$\Pt$ pions and checked by comparing the
calorimeter energy of isolated tracks in the collision data to their
momenta as measured in the COT. Further details of the CDF~II
simulation can be found in~\cite{sim}.

\subsection{\label{sec:Wplusjets} {\boldmath$W$}+Jets Dataset}
From the inclusive lepton dataset produced by the electron and muon
triggers described in Section~\ref{sec:CollBeam}, we select events
with an isolated electron $\Et$ (muon $\Pt$) greater than $20~\GeV$
and $\met>20~\GeV$. The isolation $I$ of the electron or muon is
defined as the calorimeter transverse energy in a cone of $\Delta
R\equiv\sqrt{\Delta\eta^2+\Delta\phi^2}<0.4$ around the lepton (not
including the lepton energy itself) divided by the $\Et$ ($\Pt$) of
the lepton.  We require $I<0.1$.  The $W$+jets dataset is
categorized according to the number of jets with $\Et>15~\GeV$ and
$|\eta|<2.0$. The decay of $\ttbar$ pairs gives rise to events with
typically at least three such jets, while the $W$ plus one or two
jet samples provide a control dataset with little signal
contamination. Jets are identified using a fixed-cone algorithm with
a cone size of 0.4 and are constrained to originate from the
$\ppbar$ collision vertex. Their energies are corrected to account
for detector response variations in $\eta$, calorimeter gain
instability,  and multiple interactions in an event. A complete
description of $W$+jets event selection is given in~\cite{kinPRD}.

\begin{figure}[htbp]
\begin{center}
\includegraphics[width=4.0in]{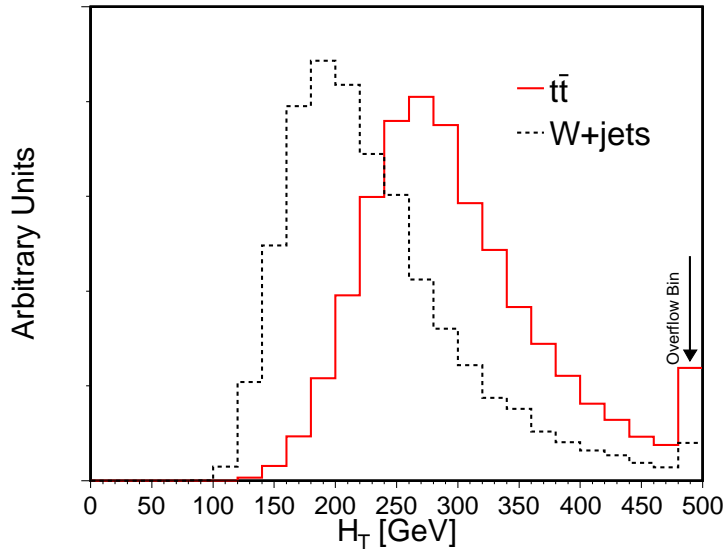}
\caption{$\Ht$ distributions, normalized to unity, for $t\bar t$
(solid line) and $W$+jets (dotted line) {\tt PYTHIA} Monte Carlo
events with three or more jets after the event selection described
in the text.} \label{fig:Ht}
\end{center}
\end{figure}

The $W$+jets dataset consists mainly of events of direct production
of $W$ bosons with multiple jets.  This amounts also to the largest
background to $\ttbar$ signal. As a first stage of background
reduction, we define a total event energy, $\Ht$, as the scalar sum
of the electron $\Et$ or muon $\Pt$,  the event $\met$ and jet $\Et$
for jets with $\Et>8~\GeV$ and $|\eta|<2.4$. Due to the large mass
of the top quark, a $\ttbar$ event is expected to have a large $\Ht$
compared to a $W$ plus three or more jets event, as illustrated in
Figure~\ref{fig:Ht}. We studied the expected amount of signal (S)
and background (B) as a function of $\Ht$ using the {\tt PYTHIA}
Monte Carlo program to model the signal $\Ht$ distribution. Data is
used to model the background $\Ht$ distribution. We optimized the
selection of events by imposing a minimum $\Ht$ requirement which
maximizes $S/\sqrt{S+B}$. We select events with $\Ht>200~\GeV$,
rejecting approximately 40\% of the background while retaining more
than 95\% of the $\ttbar$ signal.

There are 337 $W$ plus three or more jet events with $\Ht>200~\GeV$
in 194 pb$^{-1}$ of data, 115 from $W\rightarrow\mu\nu$ candidates
and 222 from $W\rightarrow e\nu$ candidates.

Even after the $\Ht$ cut, the expected $S\!:\!B$ in $W$ plus three
or more jet events is only of order 1:3. To further improve the
signal to background ratio, we identify events with one or more
$b$-jets by searching inside jets for semileptonic decays of $b$
hadrons into muons.

\subsection{\label{sec:SLT}Soft Lepton Tagging Algorithm}
Muon identification at CDF proceeds by extrapolating tracks found in
the central tracker, through the calorimeter to the muon chambers,
and matching them to track segments reconstructed in the muon
chambers. Matching is done in the following observables:
extrapolated position along the muon chamber drift direction ($x$),
the longitudinal coordinate along the chamber wires ($z$) when such
information is available, and the extrapolated slope compared to the
slope of the reconstructed muon chamber track segment ($\phi_L$).
Tracks are paired with muon chamber track segments based on the best
match in $x$ for those track segments that are within 50~cm of an
extrapolated COT track. In what follows we refer to the difference
between the extrapolated and measured positions in $x$ and $z$ as
d$x$ and d$z$, respectively, and the extrapolated and measured slope
as d$\phi_L$. The distributions of these variables over an ensemble
of events are referred to as the matching distributions. In addition
to selection based on d$x$ and d$z$, the standard muon
identification also requires consistency with minimum ionizing
energy deposition in the calorimeters. However, in order to retain
sensitivity for muons embedded in jets, the muon SLT algorithm makes
full usage of the muon matching information without any requirement
on the calorimeter information. The algorithm starts with
high-quality reconstructed tracks in the COT, selected by requiring
at least 24 axial and 24 stereo hits on the track. Some rejection
for pion and kaon decays in flight is achieved by requiring that the
impact parameter of the reconstructed track be less than 3~mm with
respect to the beamline. The track is also required to originate
within 60~cm in $z$ of the center of the detector. Only tracks
passing these cuts and extrapolating within $3\sigma(\Pt)$ in $x$
outside of the muon chambers, where $\sigma(\Pt)$ is the multiple
scattering width, are considered as muon candidates. Also, when a
track extrapolates to greater than $3\sigma(\Pt)$ in $x$ inside the
muon chambers, but no muon chamber track segment is found, the track
is rejected and not allowed to be paired to other muon chamber track
segments.

Candidate muons are selected with the SLT algorithm by constructing a quantity
$L$, based on a comparison of the measured matching variables with
their expectations. To construct $L$ we first form a sum, $Q$, of individual $\chi^2$
variables

\begin{equation}\label{eq:chi2}
    Q=\sum_{i=1}^n\frac{(X_i-\mu_i)^2}{\sigma_i^2},
\end{equation}
\noindent where $\mu_i$ and $\sigma_i$ are the expected mean and
width of the distribution of matching variable $X_i$. The sum is
taken over $n$ selected variables, as described below. $L$ is then
constructed by normalizing $Q$ according to

\begin{equation}\label{eq:likelihood}
    L=\frac{(Q-n)}{\sqrt{{\rm var}(Q)}},
\end{equation}

\noindent where the variance var$(Q)$ is calculated using the full
covariance matrix for the selected variables. The normalization is
chosen to make $L$ independent of the number of variables $n$; note
that the distribution of $L$ tends to a Gaussian centered at zero
and with unitary width, for $n$ sufficiently large. The correlation
coefficients between each pair of variables are measured from
$J/\psi\ra\mu\mu$ data. The calculation proceeds by comparing the
variance of the sum with the sum of the variances of each pair of
$\chi^2$ variables in Equation~\ref{eq:chi2}. Since the values of
the matching variables are either  positive or negative, according
to the local coordinate system, separate correlation coefficients
are used for pairs that have same-sign and opposite-sign values.

The selected variables are the full set of matching variables,
$x,~z,~\phi_L$ in the CMU, CMP and CMX with the following two
exceptions:  The CMP chambers do not provide a measurement of the
longitudinal coordinate $z$, and matching in $\phi_L$ is not
included for track segments in the muon chambers that have only
three hits. Because of their significantly poorer resolution, track
segments reconstructed in the CMU chambers with three hits are not
used. Note that a muon that traverses both the CMU and the CMP
chambers yields two sets of matching measurements in $x$ and
$\phi_L$ and one $z$ matching measurement, and are referred as CMUP
muons.  All available matching variables are used in the calculation
of $L$ for a given muon candidate. By placing an appropriate cut on
$L$, background is preferentially rejected because hadrons have
broader matching distributions than muons since the track segments
in the muon chambers from hadrons are generally a result of leakage
of the hadronic shower.

\begin{figure}[htbp]
\begin{center}
\includegraphics[width=5.cm]{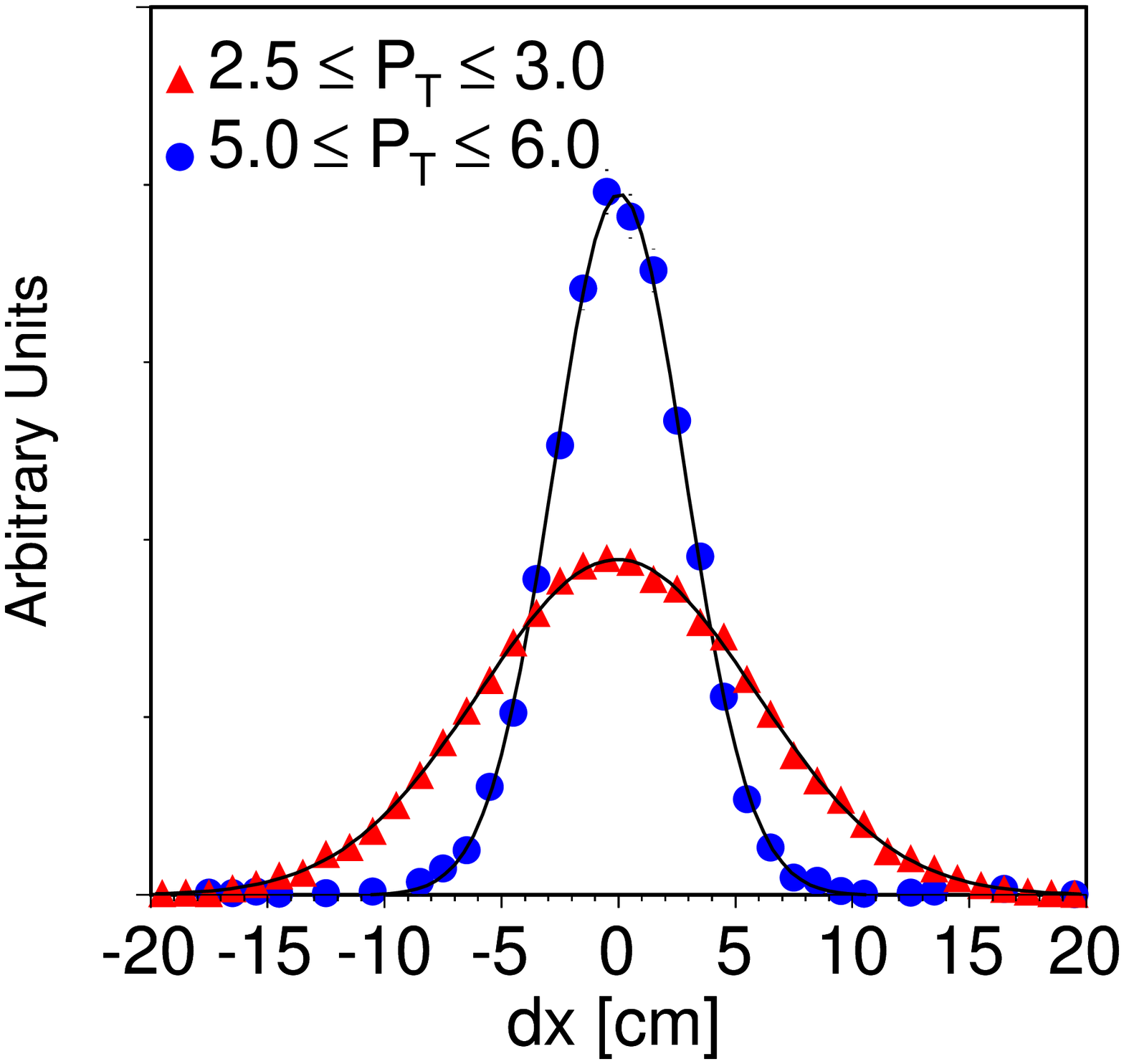}
\includegraphics[width=5.cm]{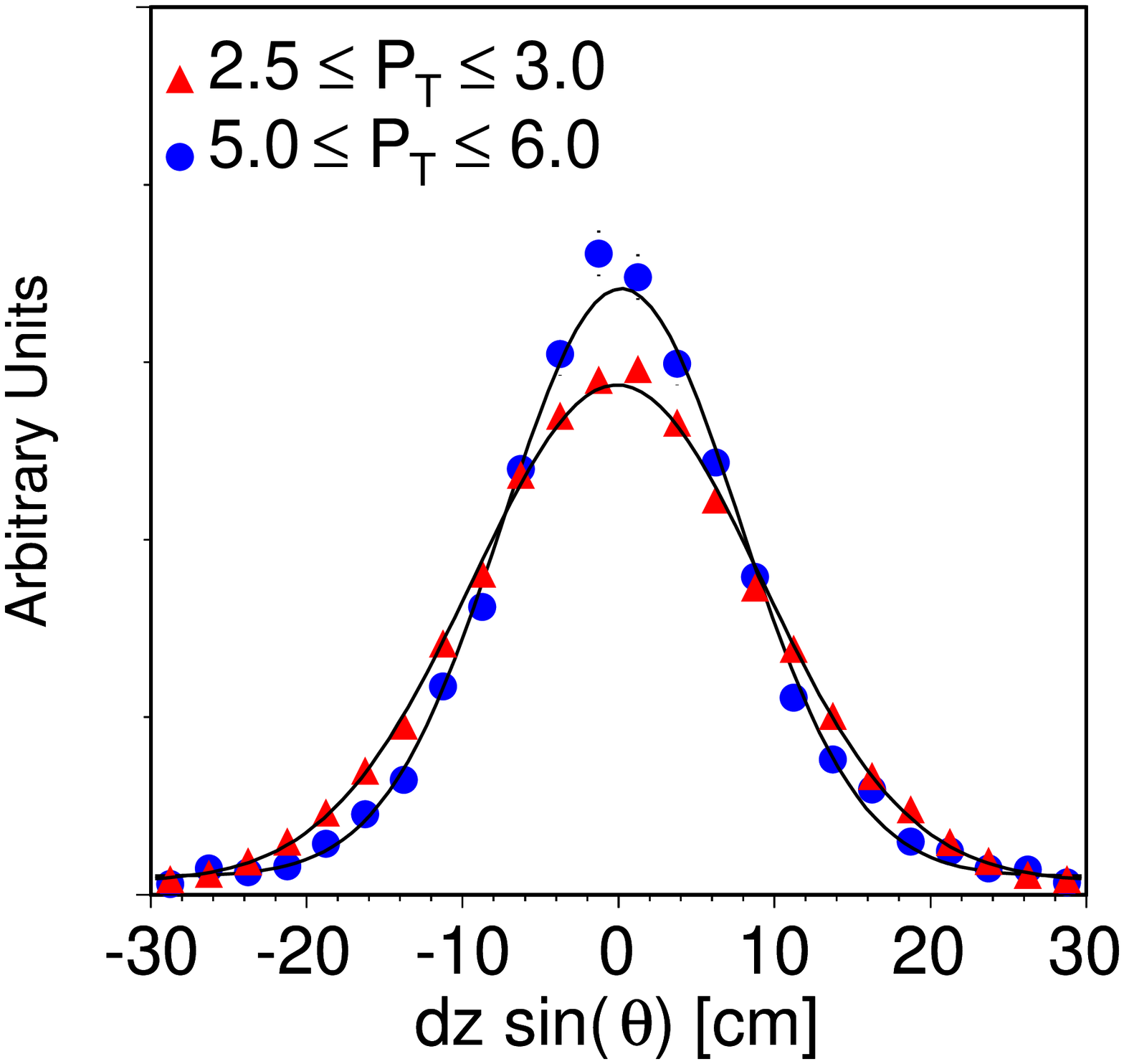}
\includegraphics[width=5.cm]{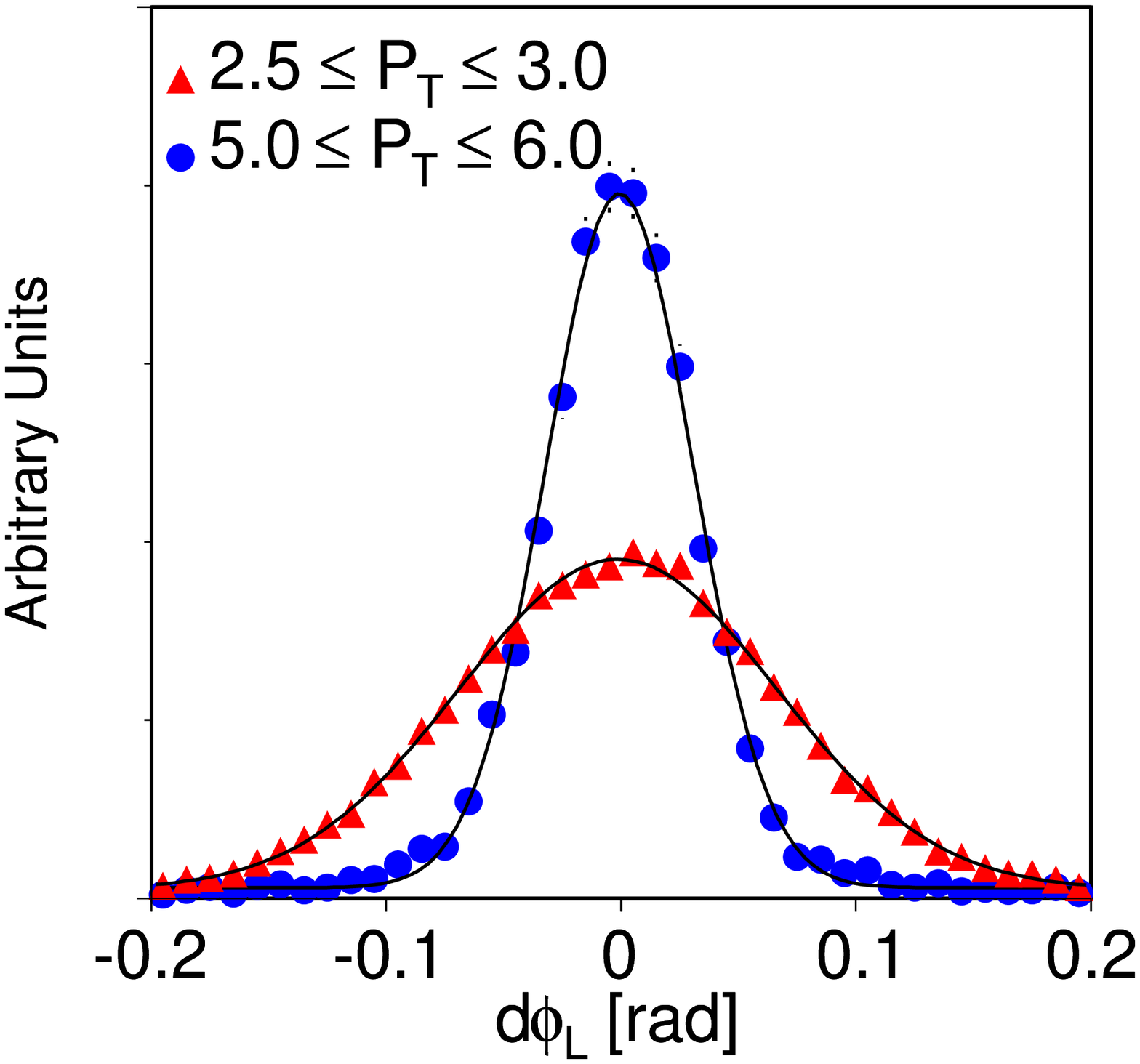}
\caption{Examples of muon matching distributions for the CMU in (left to right) drift direction, $x$,
the longitudinal coordinate, $z$, and angle, $\phi_L$. For each variable we show two $\Pt$ ranges,
$2.5\le\Pt \le 3~\GeV/c$ (wide distribution) and $5\le\Pt\le 6~\GeVc$ (narrow distribution).
For the longitudinal coordinate $z$, we plot the product of d$z$ by $\sin\theta$, which is the
projection orthogonal to the direction of flight.}
\label{fig:mumatch}
\end{center}
\end{figure}

The widths of the matching distributions that enter into $L$ are a
combination of intrinsic resolution of the muon chambers and
multiple scattering. The multiple scattering term varies inversely
with $\Pt$ and is dominant at low $\Pt$. The expected widths of the
matching distributions are based on measurements of muons from
$J/\psi$ decays at low $\Pt$ (see Figure~\ref{fig:mumatch}) and $W$
and $Z$ boson decays at high $\Pt$.

\begin{figure}[htbp]
\begin{center}
\includegraphics[width=3.0in]{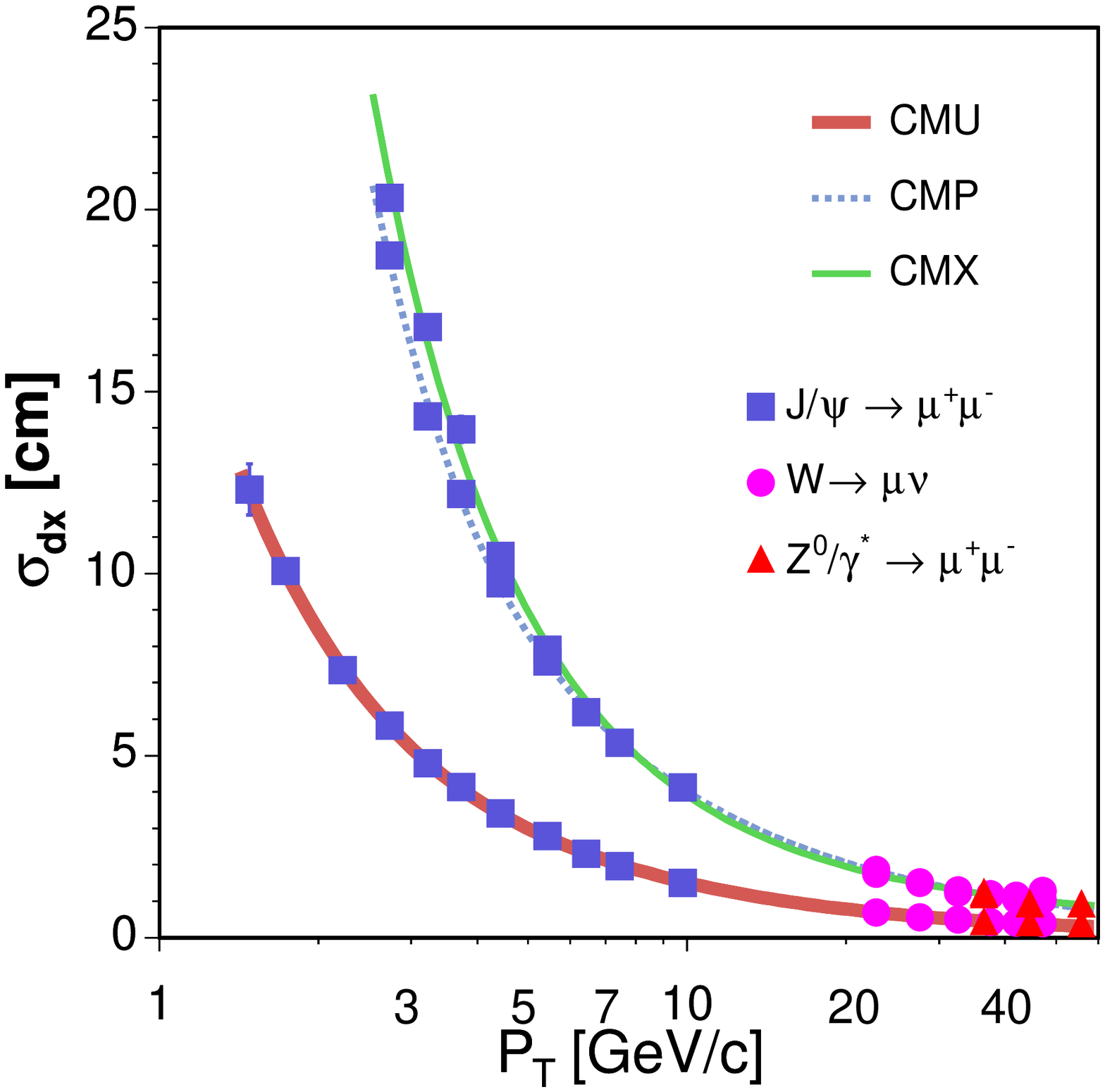}
\includegraphics[width=3.0in]{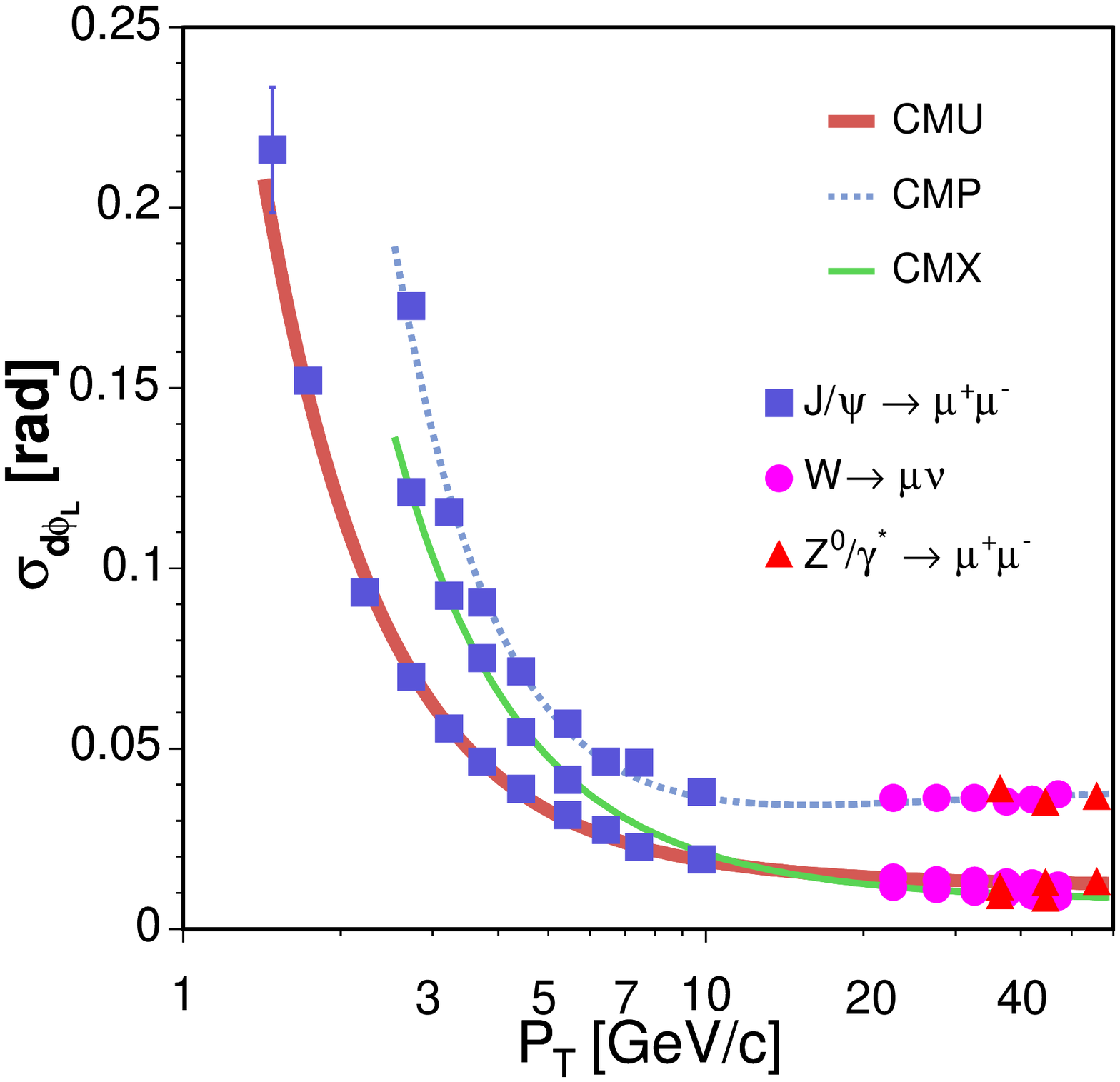}
\includegraphics[width=3.0in]{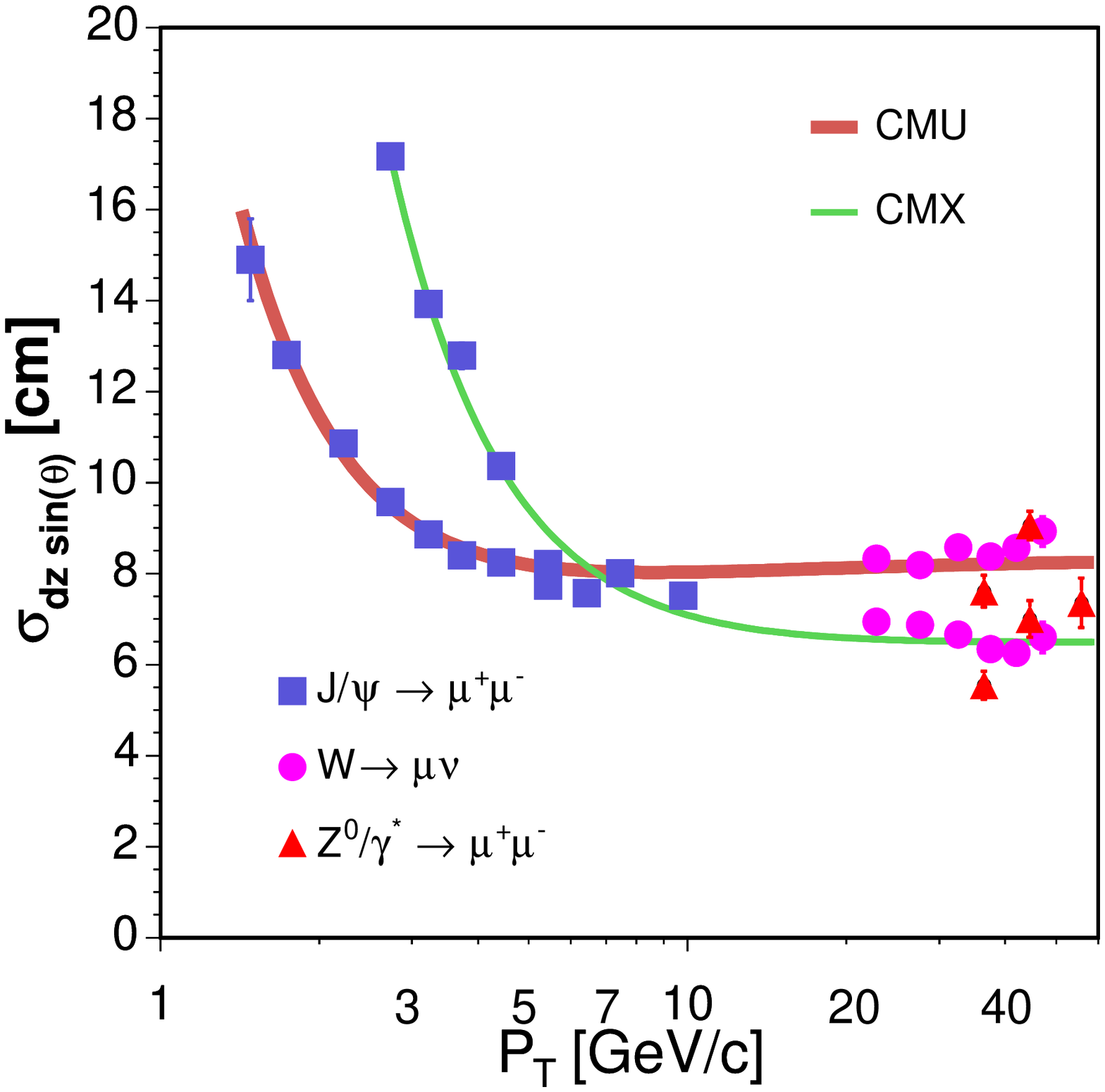}
\caption{Width of the matching variable distributions
($\mathrm{d}x,\mathrm{d}zsin\theta,\mathrm{d}\phi_L$) vs.\@ $\Pt$
for each of the muon chambers. The data points, from $J/\psi$, $W$
and $Z$ decays, are fit to the parameterizations described in the
text. } \label{fig:cmuwidth}
\end{center}
\end{figure}

The mean values ($\mu_i$ in Equation~\ref{eq:chi2}) are typically
zero, except for a small offset in the CMU d$z$.  We parameterize
the widths as a function of up to three variables: $\Pt,~\eta$ and
$\phi$. These variables describe to first order the effects of
multiple scattering in the detector.  For the CMU detector, $\Pt$ is
sufficient since the material traversed by a muon candidate is
approximately homogenous in $\eta$ and $\phi$. The widths are
parameterized with a second-order polynomial in $1/\Pt$ with an
exponential term that describes the $\Pt$ range below $3~\GeVc$. For
the CMP detector we parameterize the widths as functions of $\Pt$
and $\phi$ to take into account the rectangular shape of the
absorber outside the central calorimeter. For the CMX detector we
use $\Pt$, $\eta$ and $\phi$ to account for a number of
irregularities in the amount of absorber between $\eta=0.6$ and
$\eta=1.0$. The measurement of the widths of the matching
distributions as functions of $\Pt$, overlayed with their fits, are
shown in Figure~\ref{fig:cmuwidth}.

\begin{figure}[htbp]
\begin{center}
\includegraphics[width=7.5cm]{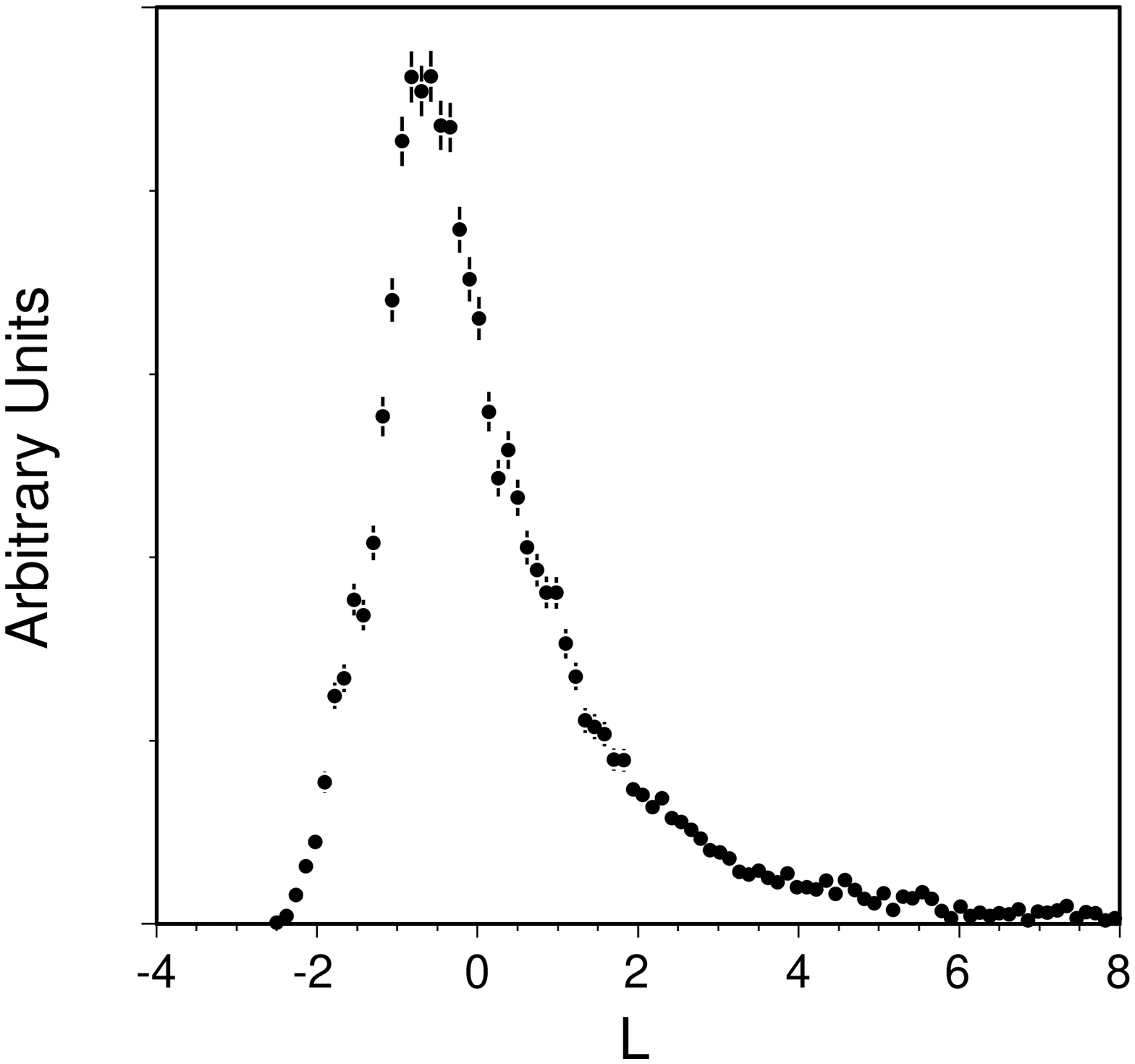}
\includegraphics[width=7.5cm]{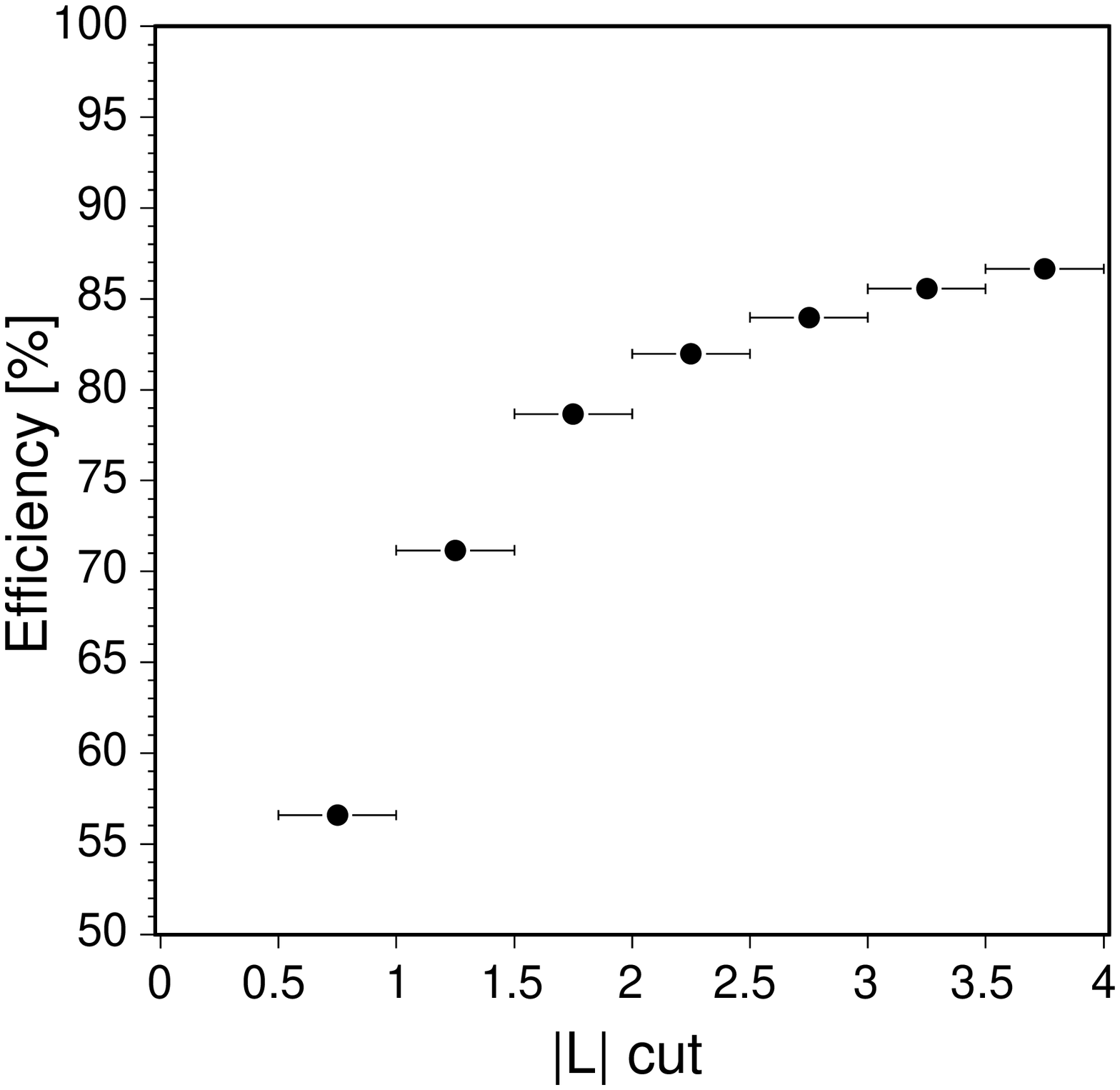}
\caption{Left: distribution of $L$ from  $J/\psi$ decays. Right: the
SLT efficiency as a function of the $|L|$ cut, as measured from
$J/\psi$ decay.} \label{fig:likeh}
\end{center}
\end{figure}

Figure~\ref{fig:likeh} (left) shows an example of the distribution
of $L$ from $J/\psi$ decays. The number of variables used varies
from two to five. Figure~\ref{fig:likeh} (right) shows the
efficiency of the SLT algorithm as a function of $L$ from $J/\psi$
data. The efficiency plateaus at about 85\% for $|L|\ge 3.5$.

\subsection{\label{sec:Evsel}Event Tagging}
\begin{figure}[htbp]
\begin{center}
\includegraphics[width=4.0in]{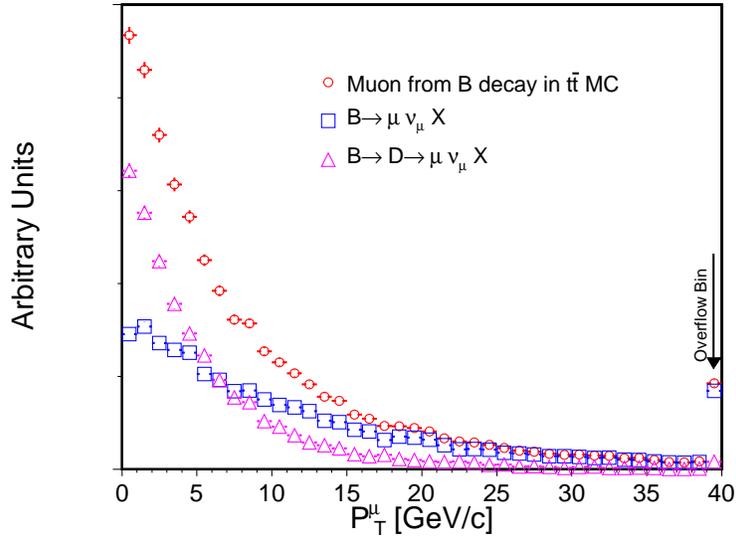}
\caption{$\Pt$ distribution of muons from $b$ hadron decays in {\sc
PYTHIA} Monte Carlo top events.  The circles are all muons from $b$
hadron decays.  The triangles are direct $B\rightarrow\mu\nu X$
decays and the squares are sequential $B\rightarrow
D\rightarrow\mu\nu X$ (where ``B" indicates a $b$ hadron).}
\label{fig:pythiamupt}
\end{center}
\end{figure}

In this analysis we seek to identify semileptonic decays of $b$
hadrons inside jets in $\ttbar$ events.  The transverse momentum
spectrum of these muons, covering a broad range from a few $\GeVc$
to over $40~\GeVc$, is shown in Figure~\ref{fig:pythiamupt} from the
{\tt PYTHIA} Monte Carlo sample.  Within the $W$+jets dataset
defined in Section~\ref{sec:Wplusjets}, we isolate a subset of
events with at least one ``taggable" track. A taggable track is
defined as any track, distinct from the primary lepton, passing the
track quality requirements described in Section~\ref{sec:SLT}, with
$\Pt>3~\GeVc$, within $\Delta R<0.6$ of a jet axis and pointing to
the muon chambers to within a 3$\sigma$ multiple scattering window
(the $\sigma$ of the multiple scattering window is defined as the
$\sigma_{\mathrm{d}x}$ shown in Figure~\ref{fig:mumatch}). The
$z$-coordinate of the track at the origin must be within 5~cm of the
reconstructed event vertex (the vertex reconstruction is described
in detail in~\cite{SVX}). Jets are considered ``SLT tagged" if they
contain a taggable track, which is also attached to a track segment
in the muon chambers and the resulting muon candidate has $|L|<3.5$.

A potentially large background arises from $J/\psi$ decay and
sequential, double semi-leptonic $b\rightarrow c\rightarrow s$
decay, resulting in one lepton from the $b$ decay and an oppositely
charged lepton from the $c$ decay. Therefore, events are rejected if
the primary lepton is of opposite charge to a SLT muon tag and the
invariant mass of the pair is less than $5~\GeVcc$. Similarly,
events are also rejected if the primary lepton is a muon that
together with an oppositely-charged SLT muon tag forms an invariant
mass between 8 and $11~\GeVcc$ or 70 and $110~\GeVcc$, consistent
with an $\Upsilon$ or a $Z$ particle, respectively. The sequential
decay cut and the $\Upsilon$ and $Z$ removal reduce the $\ttbar$
acceptance by less than 1\%.

Events passing all event selection cuts that have at least one
taggable track are referred to as the `pretag' sample.  There are
319 pretag events with three or more jets, 211 in which the primary
lepton is an electron and 108 in which it is a muon.  Out of these
events we find 20 events with a SLT tag, 15 in which the primary
lepton is an electron and 5 in which it is a muon. This set of
events is the $\ttbar$ candidate sample from which we measure the
$\ttbar$ production cross section in Section~\ref{sec:results}.

\section{\label{sec:Bak}Backgrounds}
In this section we describe the evaluation of background events in
the $\ttbar$ candidates sample.  The background contributions are
mostly evaluated directly from the data.  The dominant background in
this analysis is from $W$ plus jets events where one jet produces an
SLT tag. The estimate of this background relies on our ability to
predict the number of such SLT tags starting from the pretag sample.
The prediction is based on the probability for a given track in a
jet to yield an SLT tag, and is measured in $\gamma$+jets events. We
then evaluate the systematic uncertainty on the $W$+jets background
estimate by testing the predictive power of the measured
probabilities in a variety of data samples.  The $W$+jets background
evaluation and its systematic uncertainty is described in
Section~\ref{sec:Whf}. After $W$+jets production, the next largest
background is due to QCD multi-jet events.  The evaluation of the
QCD multi-jet contribution also relies on tagging probabilities
measured in $\gamma$+jets events. However, we must account for the
possible difference between the tagging probabilities for the QCD
events that populate the $\ttbar$ candidate sample because the
$\met$ often comes from a mismeasured jet and not from a neutrino.
The evaluation of the QCD background is described in
Section~\ref{sec:NonW}. An additional small source of background is
due to Drell-Yan events and is estimated from the data and described
in Section~\ref{sec:DY}. The remaining background contributions are
relatively small and are evaluated using Monte Carlo samples, as
described in Section~\ref{sec:MCBkg}.

\subsection{\label{sec:Whf}Backgrounds from {\boldmath$W$}+jets}
\begin{figure}[htbp]
\begin{center}
\includegraphics[width=6.5cm]{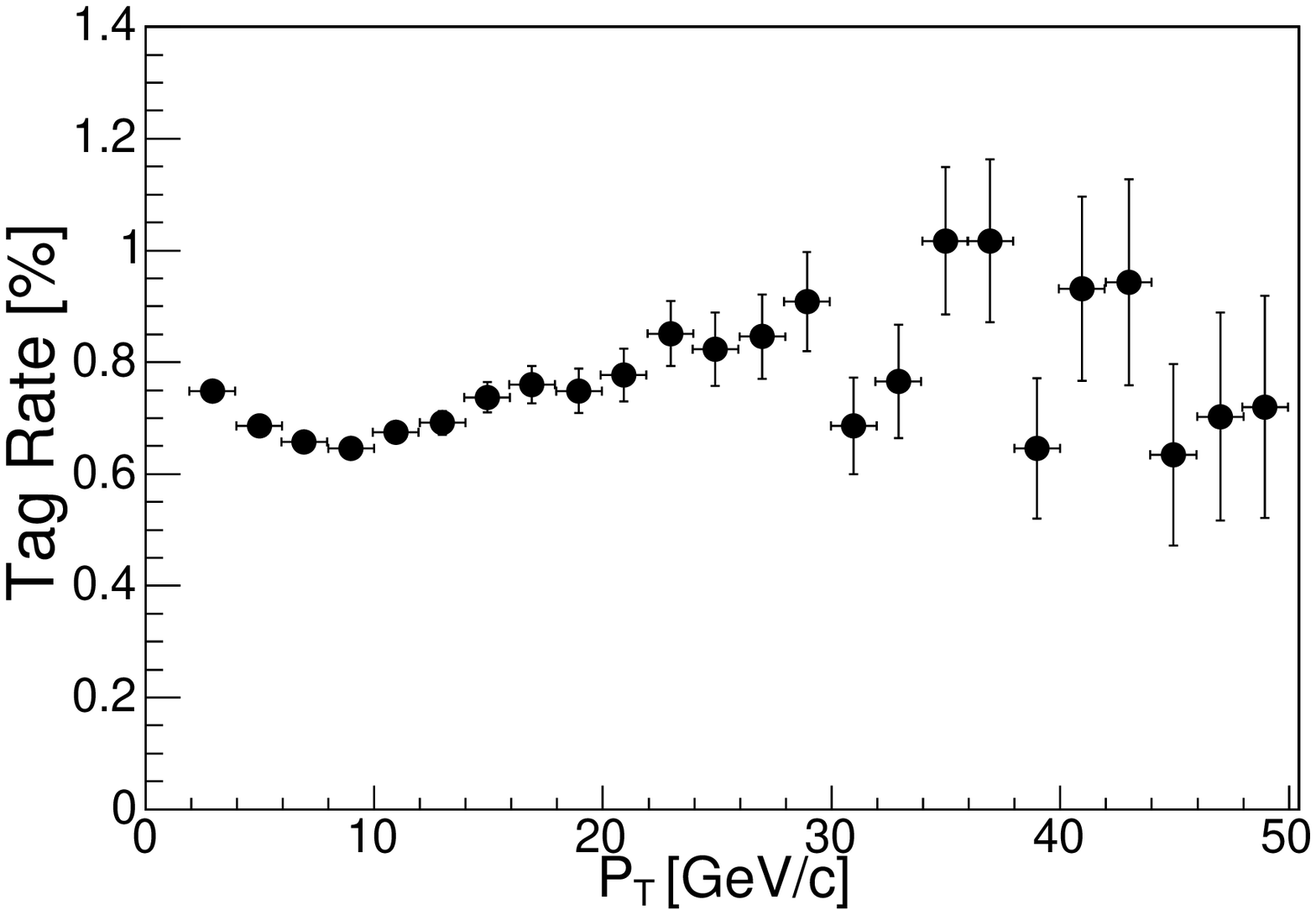}
\includegraphics[width=6.5cm]{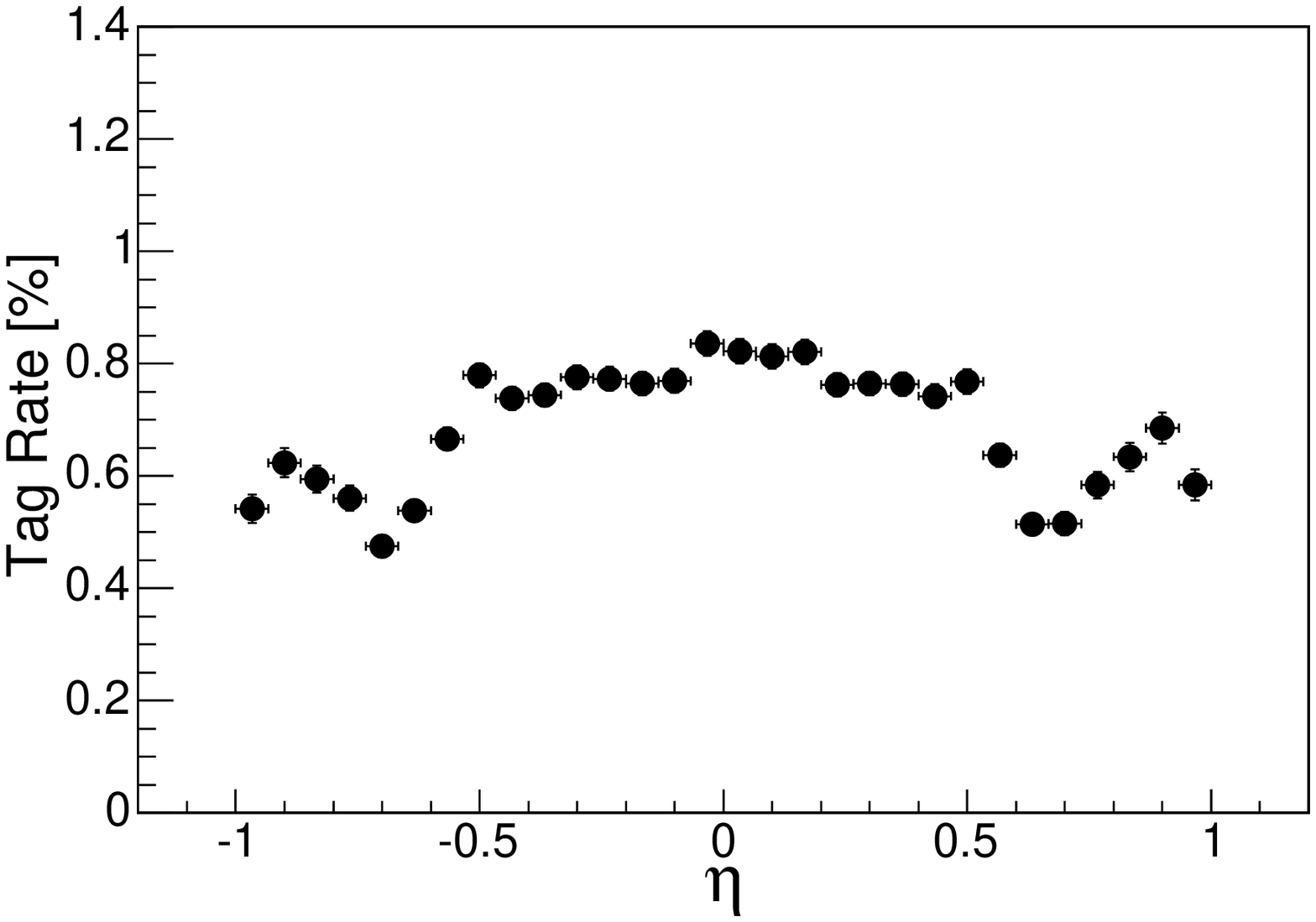}
\includegraphics[width=6.5cm]{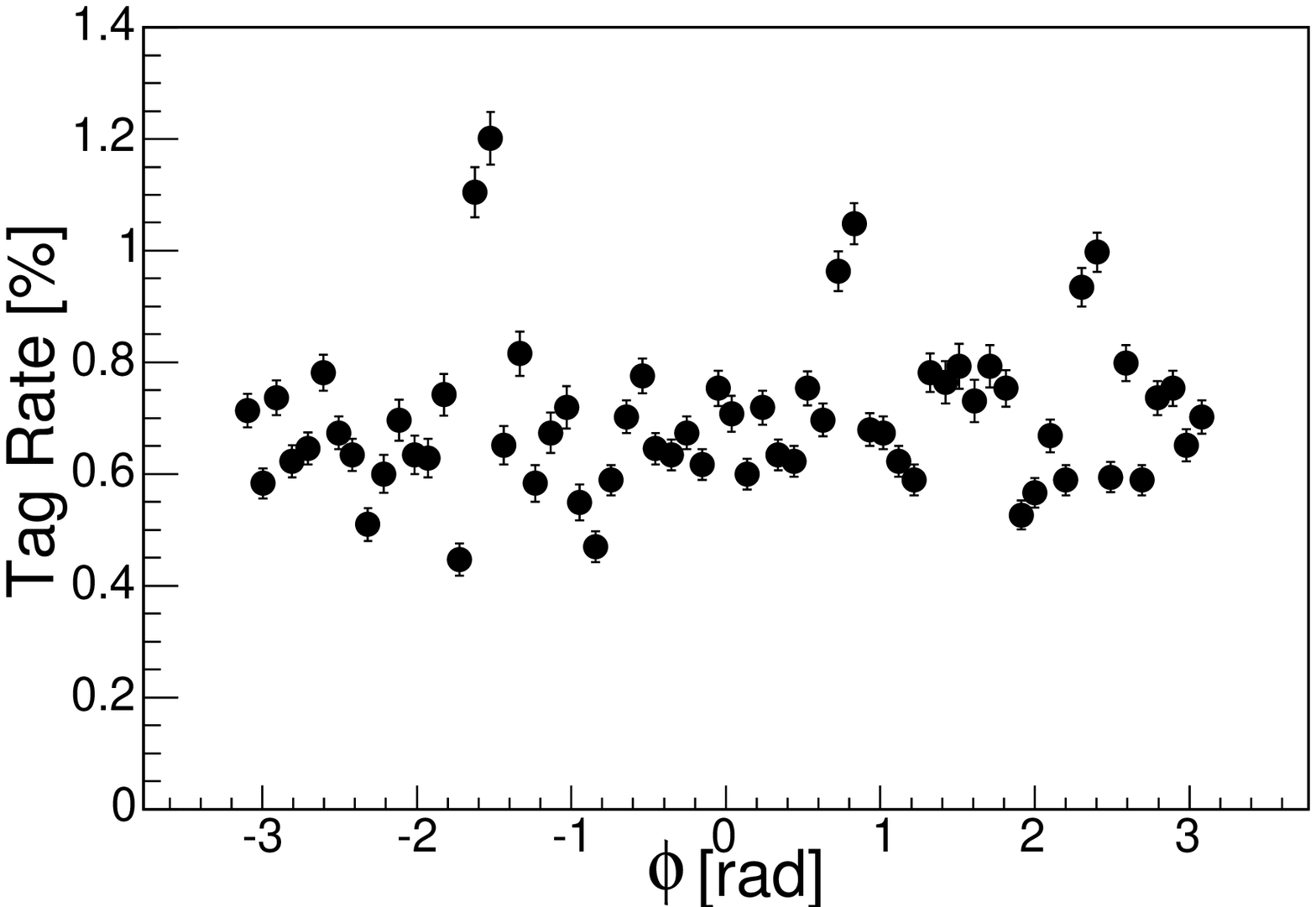}
\includegraphics[width=6.5cm]{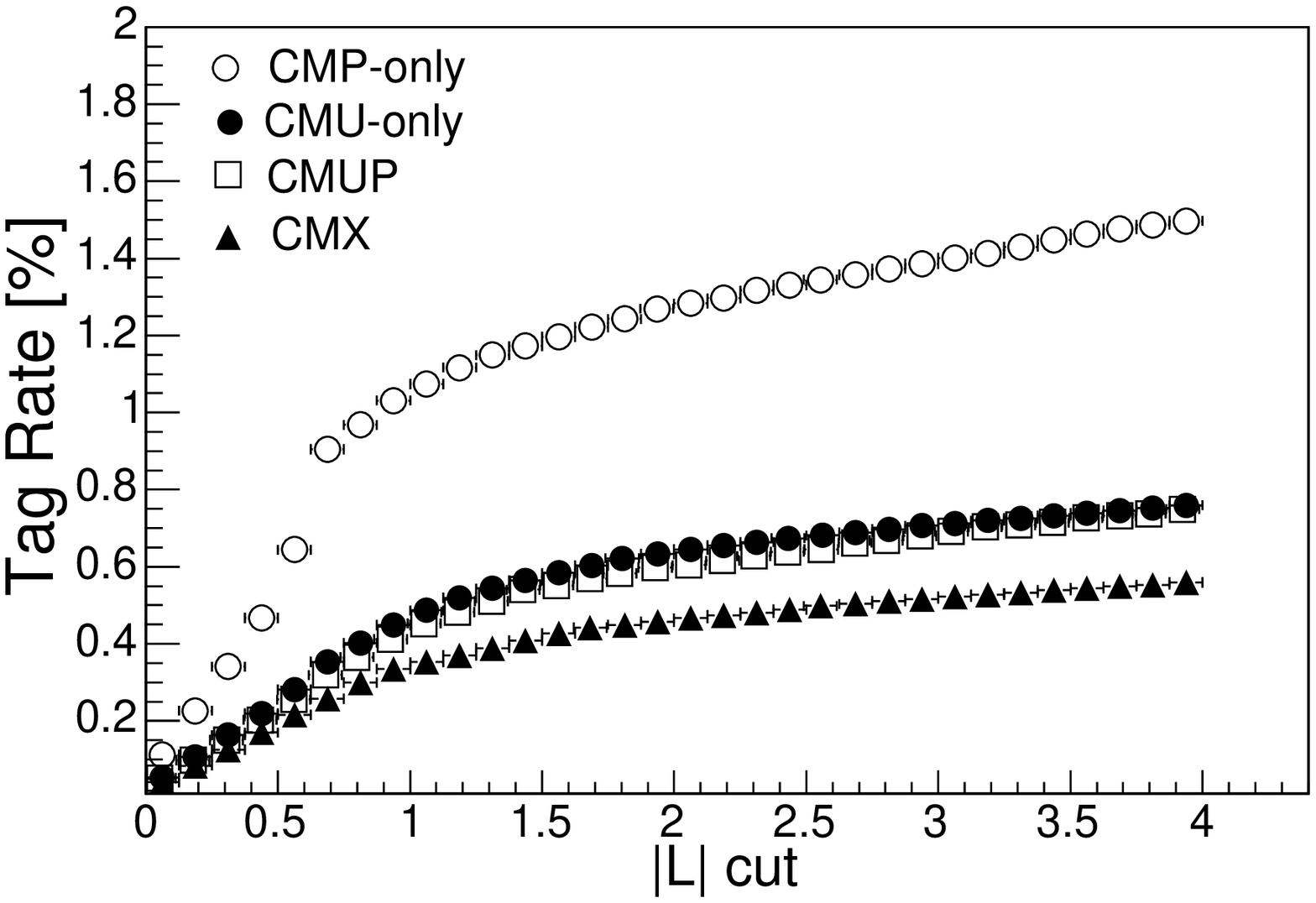}
\caption{Tag rates in the photon plus jets sample as a function of
each of the tag-rate matrix parameters.  Each plot is integrated
over the two parameters not shown. The bottom right plot shows the
tag rate as the function of the $|L|$ cut for each of the muon
categories.} \label{fig:fkrate}
\end{center}
\end{figure}

$W$ plus jets events enter the signal sample either when one of the
jets is a $b$-jet or a $c$-jet with a semileptonic decay to a muon,
or a light quark jet is misidentified as containing a semileptonic
decay (``mistagged").  We refer to these background events as
$W$+heavy flavor and $W$+``fakes", respectively. $W$+heavy flavor
events include $W\bbbar$, $W\ccbar$ and $Wc$ production.  One way of
estimating these backgrounds would be to use a Monte Carlo program,
such as {\tt ALPGEN} to predict the $W$+heavy flavor component, and
the data to predict the $W$+``fakes" (because the data provides a
more reliable measure of mistags than the simulation).  However, to
avoid double-counting, this would require an estimate of mistags
that is uncontaminated by tags from heavy flavor.  Instead we have
chosen to estimate both background components directly from the
data, and we test the accuracy of the prediction as described below.
We measure the combined $W$+heavy flavor and $W$+``fakes" background
by constructing a ``tag matrix" that parameterizes the probability
that a taggable track with a given $\Pt$, $\eta$ and $\phi$, in a
jet with $\Et>15~\GeV$, will satisfy the SLT tagging requirement
described in Section~\ref{sec:Evsel}. The variables $\eta$ and
$\phi$ are measured at the outer radius of the COT with respect to
the origin of the {CDF~II} coordinate system. The tag matrix is
constructed using jets in $\gamma$+jets events with one or more
jets. The tag probability is approximately 0.7\% per taggable track,
and includes tags from both fakes and heavy flavor. The tag rate as
a function of each of the matrix parameters (integrated over the
remaining two) is shown in Figure~\ref{fig:fkrate}. The features in
the tag rate vs.\@ $\eta$ and $\phi$ plots are a result of
calorimeter gaps and changes in the thickness of the absorber before
the muon chambers.  The matrix is binned to take account of these
variations. The bottom right plot shows the tag rate as the function
of the $|L|$ cut for each muon category. The tag rate is higher for
the CMP-only muons due to the smaller amount of absorber material
that results from cracks in the calorimeter where there is no
coverage by the CMU chambers.

We apply the tag matrix to all pretag events in the signal region
according to:

\begin{equation}
N_{\rm predicted}^{\rm tag}= \sum_{\mathrm{events}}
\left[1-\prod_{i=1}^{N_{\rm trk}}\left(1-{\cal P}_i\right)\right ] ,
\label{eq:Fake}
\end{equation}

\noindent where the sum runs over each event in the pretag sample,
and the product is over each taggable track in the event. ${\cal
P}_i$ is the probability from the tag matrix for tagging the
$i$-track with parameters $P_{T_i}$, $\eta_i$ and $\phi_i$. Note
that the sum over the events in Equation~\ref{eq:Fake} includes any
$\ttbar$ events that are in the pretag sample. We correct for the
resulting overestimate of the background at the final stage of the
cross section calculation, since the correction depends on the
measured tagging efficiency (see Section~\ref{sec:results}).

A fraction, $F_{QCD}$, of the events in the
signal region are QCD events (such as $\bbbar$ or events in which a
jet fakes an isolated lepton, see Section~\ref{sec:NonW}) for which
the background is estimated separately. Therefore, we explicitly
exclude their contribution to $N_{\rm predicted}^{\rm tag}$ and
obtain the predicted number of tagged $W$+jets background events

\begin{equation}
N_{\rm predicted}^{Wj-{\rm tag}}=(1-F_{QCD})\cdot N_{\rm predicted}^{\rm tag}.
\label{eq:bkg0}
\end{equation}

\noindent  The estimated $W$+fakes and $W$+heavy flavor background
is given in the third line of Table~\ref{tab:results}.

\begin{table}[htbp]
\caption{\label{tab:results} Number of tagged events and the
background summary. The $\Ht>200~\GeV$ requirement is made only for
events with at least 3 jets.}
 \sans
 \begin{ruledtabular}
  \begin{tabular}{ccc|cc|c}
     Background                                &  $W$ + 1 jet      & $W$ + 2 jets    &  $W$ + 3 jets  & $W$ + $\ge$ 4 jets & $W$ + $\ge$ 3 jets \\\hline
     Pre-tag Events                            &     9117          &   2170          &     211        &    108          & 319 \\
     Fake, $W b\bar b$, $W c\bar c$, $Wc$      &  116.3$\pm$11.6   & 40.5$\pm$4.1    &  7.0$\pm$0.7   & 4.3$\pm$0.4     & 11.3$\pm$1.1 \\
     $WW$, $WZ$, $ZZ$, $Z\ra \tptm$                  &   1.10$\pm$0.17   &  1.33$\pm$0.06  &  0.16$\pm$0.02 & 0.04$\pm$0.01   & 0.19$\pm$0.02 \\
     QCD                                       &   19.6$\pm$24.2   &  12.4$\pm$3.5   &  0.9$\pm$0.2   & 0.8$\pm$0.2     & 1.6$\pm$0.3 \\
     Drell-Yan                                 &    0.8$\pm$0.4    &  0.36$\pm$0.20  &  0.08$\pm$0.09 & 0.00$\pm$0.09 & 0.08$\pm$0.09\\
     Single Top                                &   0.50$\pm$0.03   &  0.94$\pm$0.06  &  0.15$\pm$0.01 & 0.035$\pm$0.003 & 0.19$\pm$0.01\\\hline
     Total Background                          &  138.2$\pm$26.8   & 55.5$\pm$5.4    &  8.2$\pm$0.8   & 5.2$\pm$0.5     & 13.4$\pm$1.3 \\
     Corrected Background                      &                   &                 &\multicolumn{2}{c|}{9.5$\pm$1.1} & 9.5$\pm$1.1 \\
     $t\bar t$ Expectation (6.7pb)             &   0.4$\pm$0.1     &  2.9$\pm$0.5    &  5.4$\pm$0.9   & 7.9$\pm$1.7     & 13.3$\pm$2.6\\\hline
     Total Background plus $t\bar t$           &  138.6$\pm$26.8   & 58.4$\pm$5.4    & \multicolumn{2}{c|}{22.8$\pm$2.8}& 22.8$\pm$2.8 \\\hline
     Tagged Events                             &  139              & 48                &    13        &       7     & 20  \\
  \end{tabular}
\end{ruledtabular}
\end{table}

The above technique relies on the assumption that the tagging rate
in jets of the $\gamma$+jets sample is a good model for the tagging
rate of the jets in $W$+jets events. The assumption is plausible
because the SLT tagging rate in generic jet events is largely due to
fakes. We have studied the heavy flavor content of the tags in the
$\gamma$+jets sample using the overlap sample between SLT tags and
displaced vertex tags identified with the silicon
tracker~\cite{SVX}. We find that (21.0$\pm$1.4)\% of the tags in the
$\gamma$+jets sample are from heavy flavor. We have used {\tt
MADEVENT}~\cite{MadEvent} to do generator-level comparisons of the
heavy-flavor fractions of $W$+jets events with those from the
$\gamma$ plus jets events that make up the tag matrix. We find that
the $\gamma$+jets sample used to make the tag matrix has
approximately 30\% more heavy flavor than the $W+\ge 3$ jet events.
Since SLT tags in $\gamma$+jets events are dominantly fakes, this
difference affects the background prediction in $W$+jets events at
only the few-percent level.

\begin{figure}[htbp]
\begin{center}
\includegraphics[width=8.5cm]{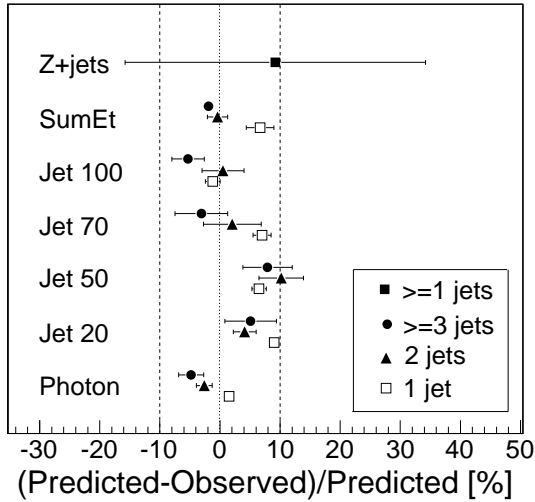}
\caption{The percent difference between the number of predicted and
measured tags in a variety of samples for different jet
multiplicities, as marked. Jet multiplicities do not include the
trigger object (photon or jet).  The three photon points contain the
events that make up the tag matrix. Their average is zero by
definition and is shown by the line centered at zero. The dashed
lines at $\pm 10$\% indicate the systematic uncertainty as
determined from these data.} \label{fig:fakesys}
\end{center}
\end{figure}

Given the limitations of a generator-level, matrix element Monte
Carlo study of the heavy flavor content of the $\gamma$+jets and
$W$+jets samples, we do not use the {\tt MADEVENT} study to evaluate
the systematic uncertainty on the background due to tagged $W$+jets
events. Instead we test the accuracy of the tag matrix for
predicting SLT muon tags by using it to predict the number of tags
in a variety of samples with different heavy flavor content. We
check $Z$ plus jets events, events triggered on a jet with $\Et$
thresholds of 20, 50, 70 and 100 $\GeV$ (called Jet~20, Jet~50,
Jet~70 and Jet~100), or triggered on four jets and the scalar sum of
transverse energy in the detector (called SumET). We find that the
matrix predicts the observed number of tags in each of these samples
to within 10\%, as shown in Figure~\ref{fig:fakesys}, and we use
this as the systematic uncertainty on the prediction from the tag
matrix.

\subsection{\label{sec:NonW}QCD Background}
We refer to events with two or more jets in which the decay of a
heavy-flavor hadron produces a high-$\Pt$ isolated lepton, or in
which a jet fakes such a lepton, as QCD events. These events enter
the sample when, in addition to the high-$\Pt$ isolated lepton, a
muon from a heavy flavor decay gives an SLT tag, or there is a fake
tag. We measure this background directly from the data.

\begin{figure}[htbp]
\begin{center}
\includegraphics[width=3.5in]{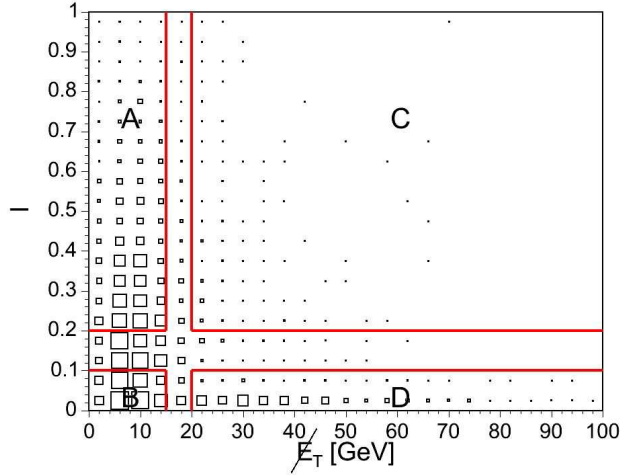}
\caption{Distribution of events with $\ge$1 jet in $\met$ vs.\@ $I$.
Regions A,B,C  are defined in the text and are used to calculate the
fraction of QCD events in region D (the signal region) according to
Equation~\ref{eq:FQCD}.} \label{fig:metvsI}
\end{center}
\end{figure}

To estimate the QCD component we first use the distribution of
pretag events in the plane of $\met$ vs.\@ isolation, $I$, of the
primary lepton. We populate this plane with lepton plus jets events
according to the event $\met$ and $I$. We consider four regions in
the plane:

\begin{eqnarray}
A: \met<15 & I>0.2 \nonumber\\
B: \met<15 & I<0.1 \nonumber\\
C: \met>20 & I>0.2 \nonumber\\
D: \met>20 & I<0.1 \nonumber
\end{eqnarray}
\noindent where Region D is the $\ttbar$ signal region.  The
distribution of events, with one or more jets, in the $\met$ vs.\@
$I$ regions is shown in Figure~\ref{fig:metvsI}.

In order to populate Regions A, B and C with only QCD events, we
correct the number of events for the expected contamination of
$W$+jets and $\ttbar$ events in those regions using expectations
from {\tt PYTHIA} $\ttbar$ and $W$ Monte Carlo simulations. The
corrections range from less than 1\% in electron plus one-jet events
in Region A, to (57$\pm$15)\% in Region B in muon plus three or more
jet events.

Assuming that the variables $\met$ and $I$ are uncorrelated for the
QCD background, the ratio of the number of QCD events in Region A to
those in Region B should be the same as the ratio of the number of
QCD events in Region C to those in Region D.  Therefore we calculate
the fraction of QCD events in Region D, $F_{QCD}$, as:

\begin{equation}
F_{QCD}=\left.{\frac{N_D^{QCD}}{N_D}}\right|_{\rm
pretag}=\left.{\frac{N_B\cdot N_C}{N_A\cdot N_D}}\right|_{\rm
pretag}, \label{eq:FQCD}
\end{equation}

\noindent where $N_D^{QCD}$ is the total number of pretag QCD events
in the signal region, and $N_i$ represent the number of events in
region $i$. The measured fractions are shown in Table~\ref{tab:Fnw}.

\begin{figure}[htbp]
\begin{center}
\includegraphics[width=4.0in]{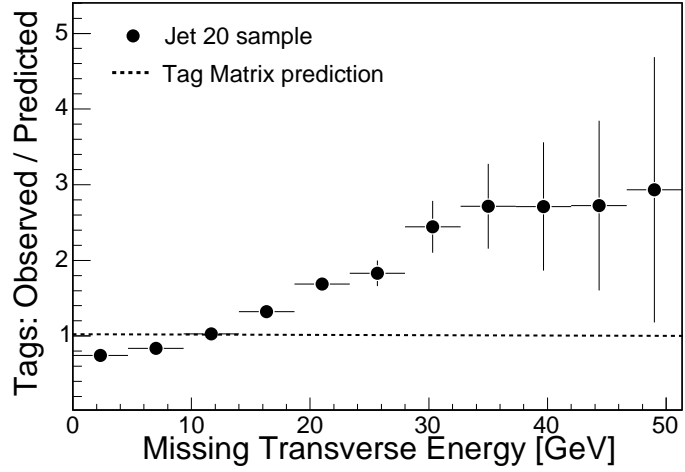}
\caption{The ratio of the number of observed tags to tags predicted
using the tag matrix, as a function of $\met$, in events with at least
one jet with measured energy above $20~\GeV$.} \label{fig:Kfactor}
\end{center}
\end{figure}

To estimate the number of tagged QCD events in the signal region, we
multiply $F_{QCD}$ by the tagging probability for QCD events.
However, this tagging probability is not necessarily given by the
tag matrix probabilities which are designed for jets in $W$+jets
events. Mismeasurement in the jet energies and differences in
kinematics between $W$+jets and QCD events may affect the tagging
probabilities. $W$+jets events have $\met$ from the undetected
neutrino, whereas QCD events have $\met$ primarily from jet
mismeasurement. Jet mismeasurement is correlated with fake tags due
to energy leakage from the calorimeter through calorimeter gaps or
incomplete absorption of the hadronic shower, both of which can
result in track segments in the muon chambers. $W$+jets events have
a primary lepton from the $W$ decay, whereas QCD events have a
primary lepton that is either a fake or a result of a semileptonic
decay of heavy flavor. The presence of a lepton from heavy flavor
decay typically enhances the tag rate. Figure~\ref{fig:Kfactor}
shows the ratio of the number of measured tags in the Jet~20 sample
to the number of tags predicted by the tag matrix as a function of
$\met$.  As expected, in QCD events with large $\met$ we find a tag
rate significantly larger than that described by the tag matrix.

\begin{figure}[htbp]
\begin{center}
\includegraphics[width=4.0in]{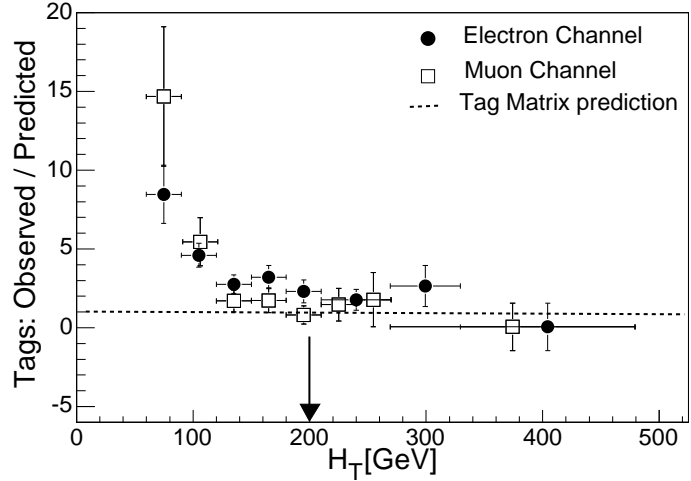}
\caption{The ratio of the observed rate of tags to that predicted,
as a function of $\Ht$ in region C for events with one or more jets.
The arrow at $200~\GeV$ shows where the selection cut for the
$\ttbar$ signal sample is placed.} \label{fig:KvsHt}
\end{center}
\end{figure}

\begin{figure}[htbp]
\begin{center}
\includegraphics[width=5.cm]{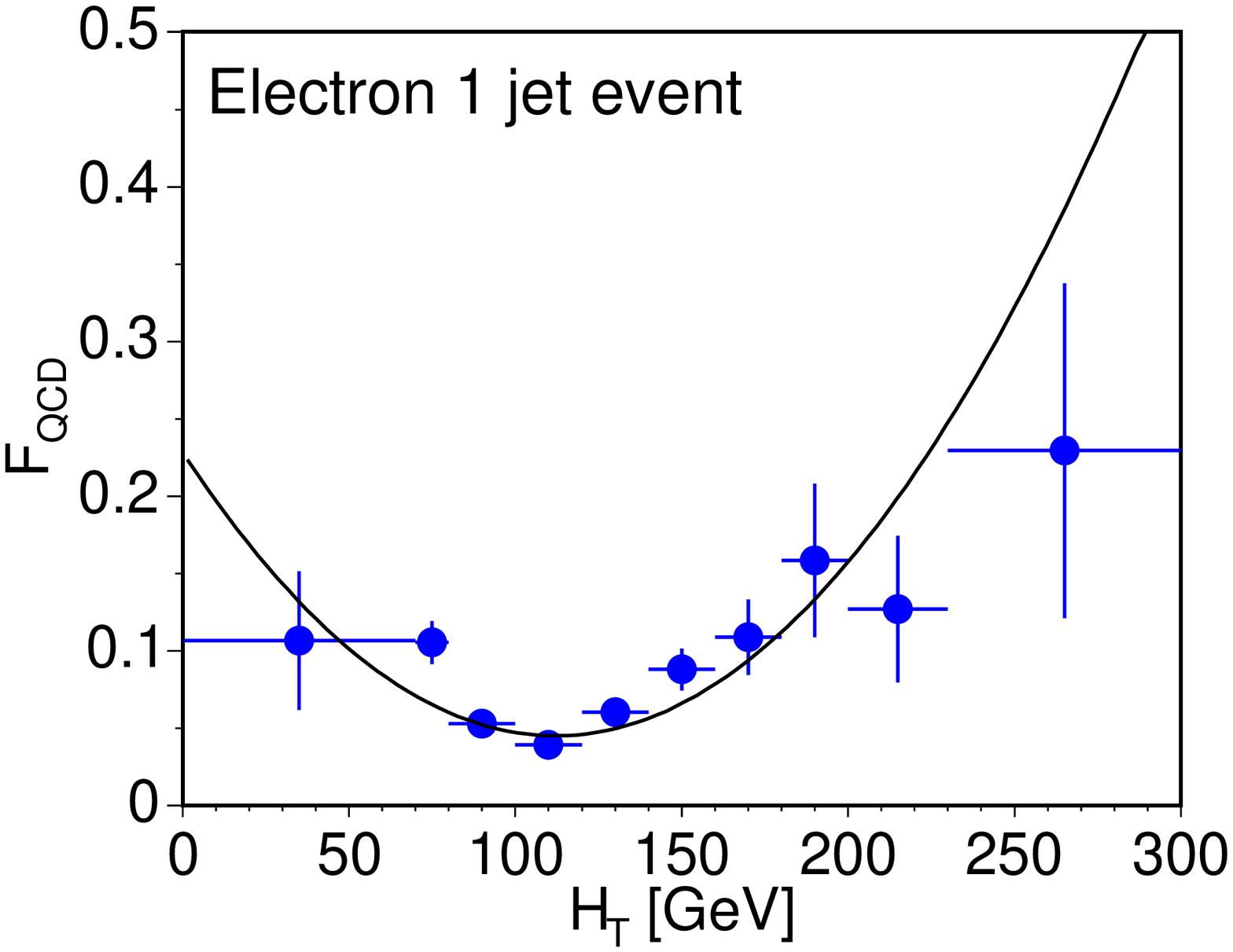}
\includegraphics[width=5.cm]{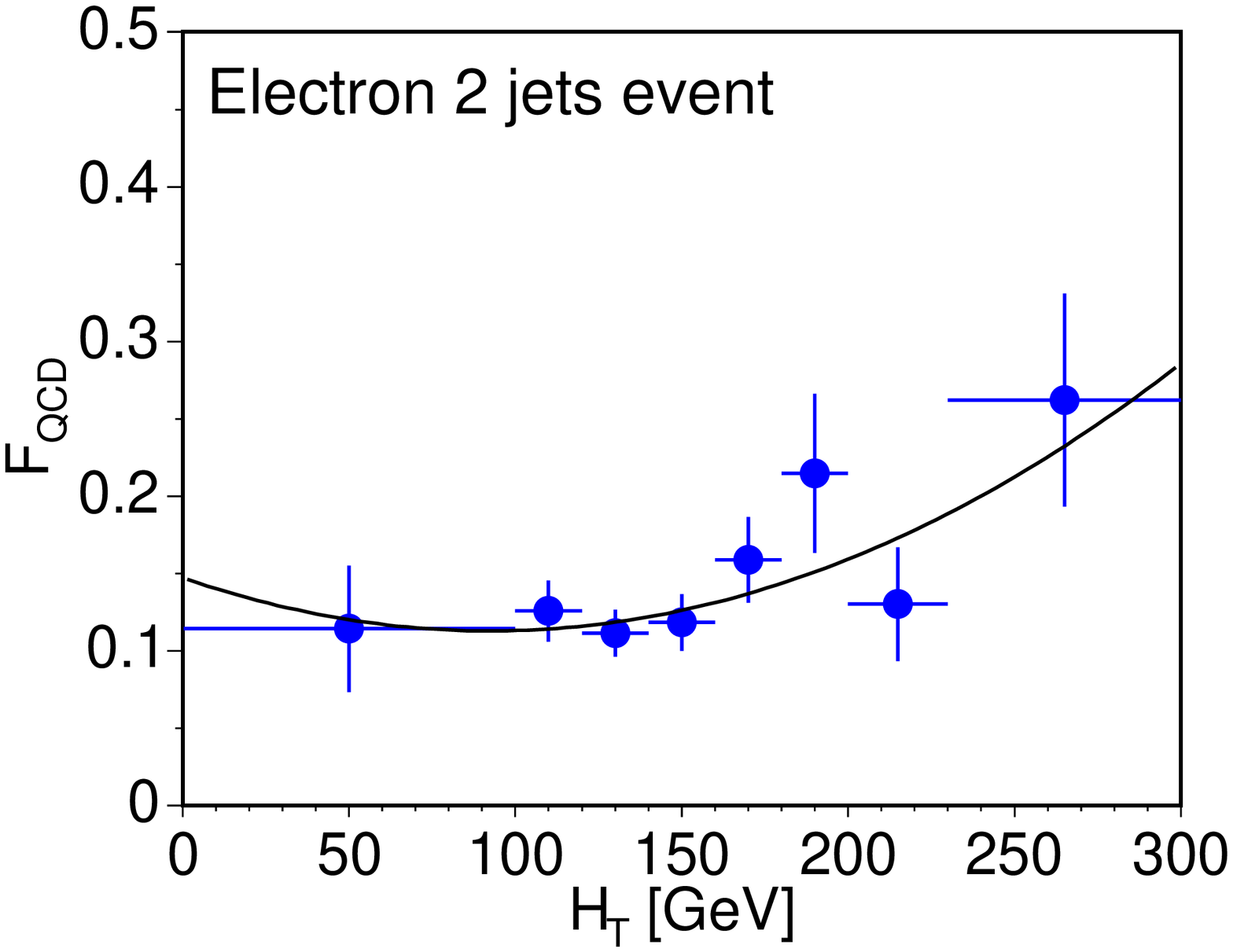}
\includegraphics[width=5.cm]{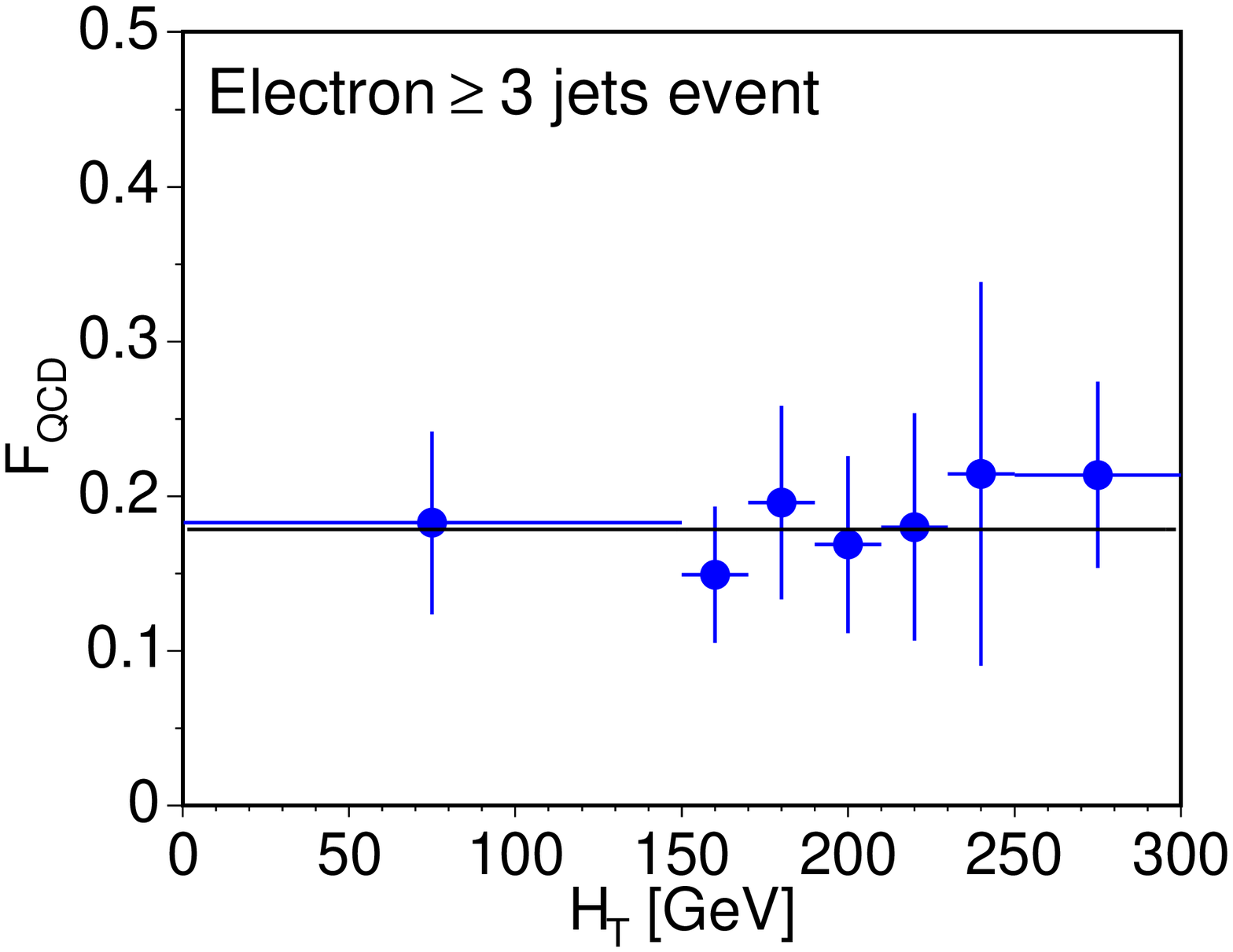}
\includegraphics[width=5.cm]{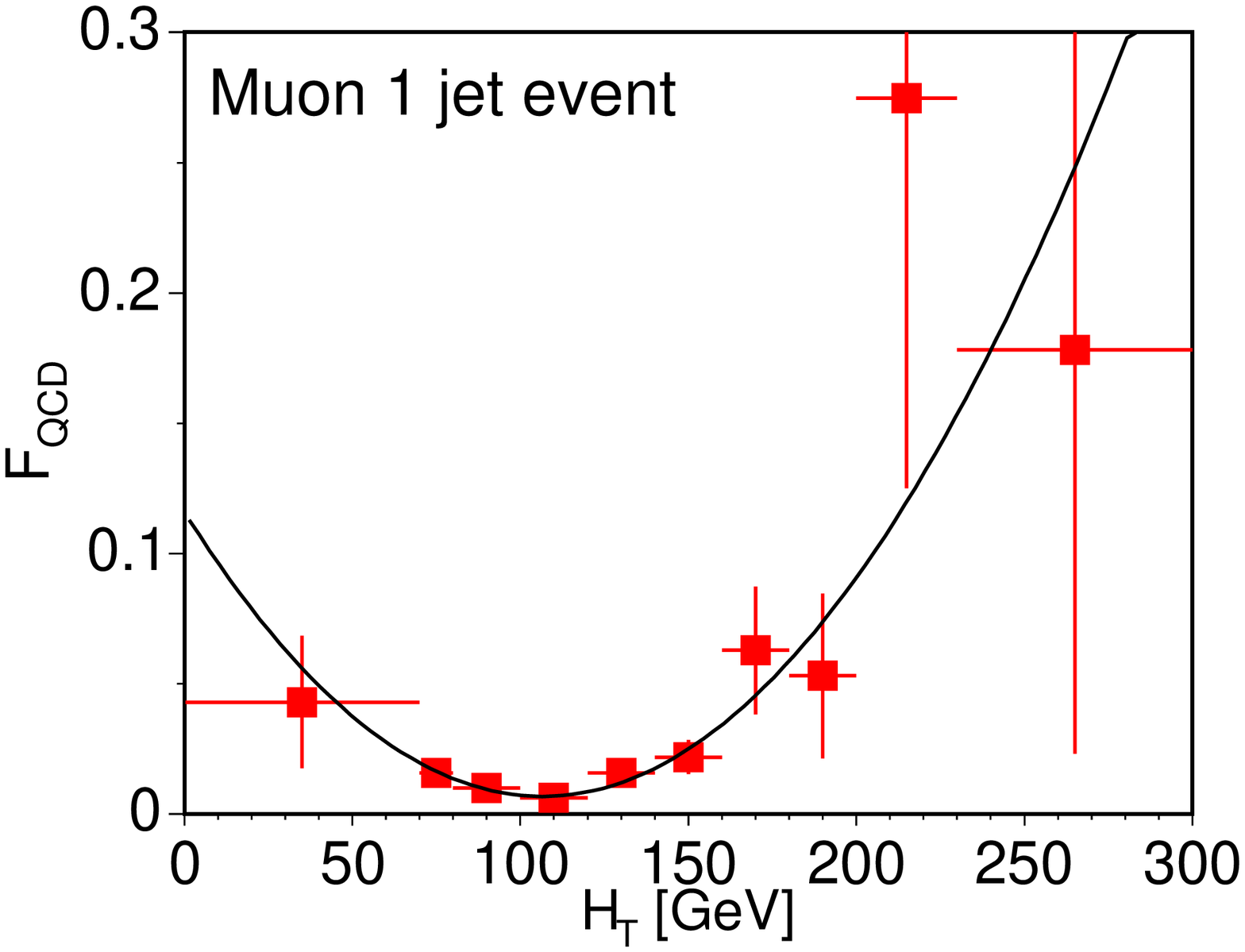}
\includegraphics[width=5.cm]{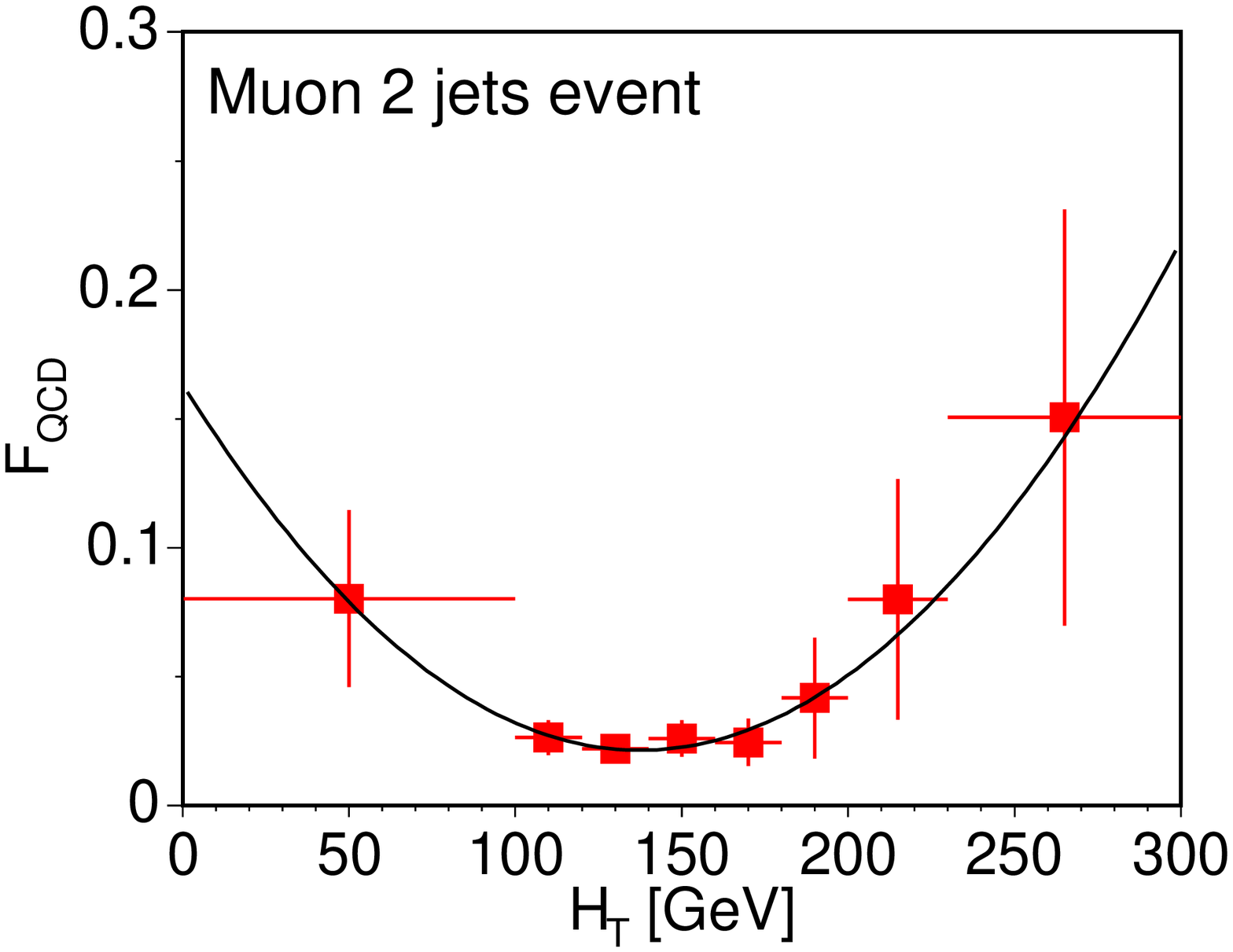}
\includegraphics[width=5.cm]{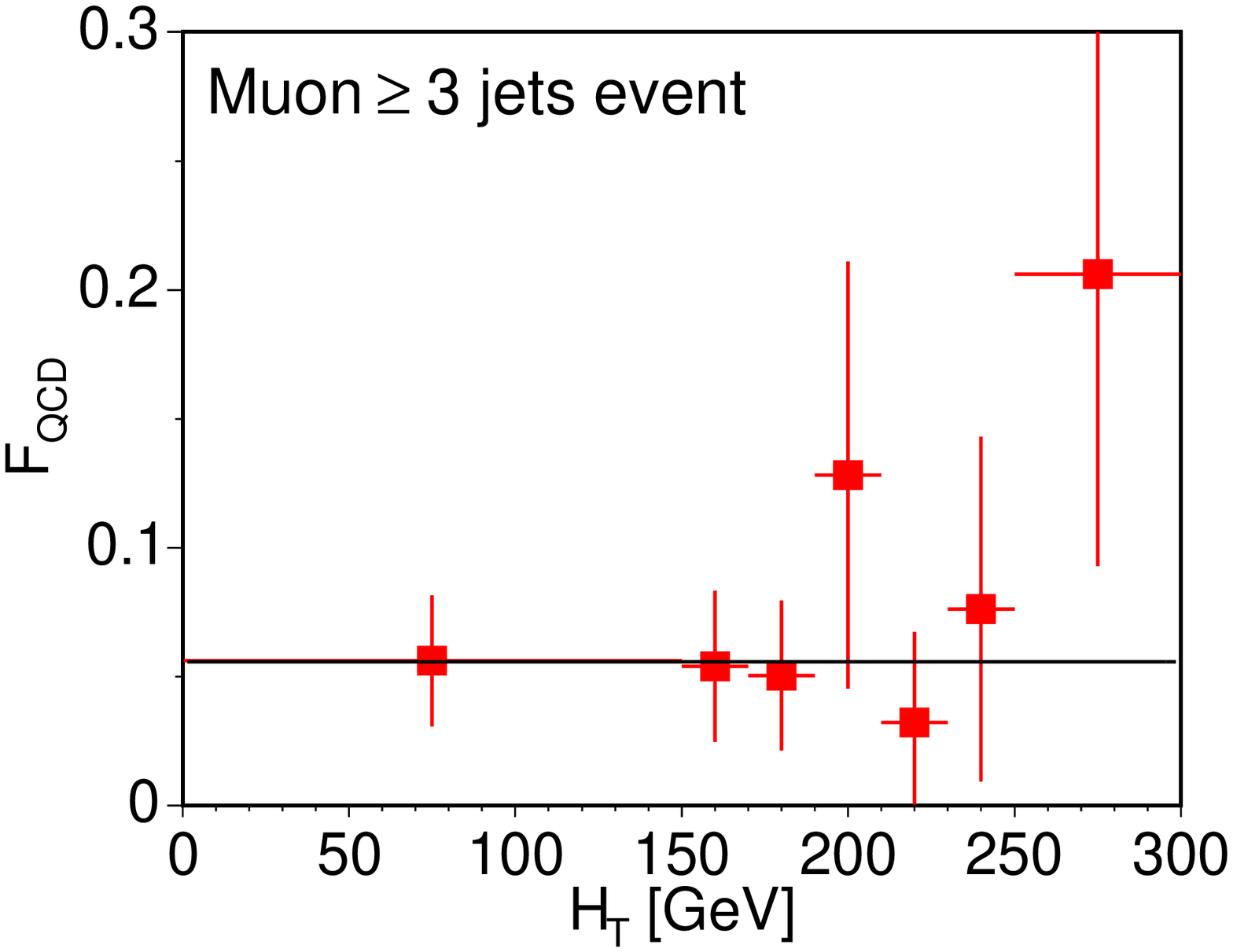}
\caption{$F_{QCD}$ measured according to Equation~\ref{eq:FQCD} as a
function of $\Ht$ in electron events (top) and muon events (bottom)
with (left to right) 1, 2 or $\ge$ 3 jets.} \label{fig:FQCDvsHt}
\end{center}
\end{figure}

We find that the prediction of the tag matrix can be renormalized to
properly account for the tag rates in QCD events with a single
multiplicative factor, which we call $k$. We measure $k$ using
events in region C by comparing the number of SLT tags found to the
number predicted by the tag matrix.  Since the signal region
contains only isolated ($I<0.1$) primary leptons, we reject events
in the measurement of $k$ in which the SLT tag is within $\Delta
R<0.5$ of the primary lepton. After this requirement we do not find
any dependence of $k$ on the isolation of the primary lepton.
Figure~\ref{fig:KvsHt} shows the ratio of measured to predicted tags
in events in region C as a function of $\Ht$. The tag rate above
$\Ht=200~\GeV$ is approximately flat and is not much different from
the prediction of the tag matrix (dashed line in
Figure~\ref{fig:KvsHt}). However, QCD events at lower $\Ht$ have a
significantly different tag rate than that predicted by the tag
matrix. As shown in Figure~\ref{fig:FQCDvsHt}, $F_{QCD}$ also has an
$\Ht$ dependence for events with 1 or 2 jets, but is flat within the
statistical uncertainty for three or more jets.

The number of QCD background events is calculated as:
\begin{equation}\label{eq:bkg1}
    N_{QCD}=\langle F_{QCD}\cdot k\rangle\cdot N_{\rm predicted}^{\rm tag},
\end{equation}

\noindent where $N_{\rm predicted}^{\rm tag}$ is given in
Equation~\ref{eq:Fake} and the brackets represent the product of
$F_{QCD}$ and $k$ convoluted with the $\Ht$ distribution of QCD
events from region C. In the control region (1 and 2 jets), the fits
of $F_{QCD}$ vs.\@ $\Ht$, shown in Figure~\ref{fig:FQCDvsHt}, are
convoluted with $k$ vs.\@ $\Ht$, shown in Figure~\ref{fig:KvsHt}.
For events with three or more jets, since there is no visible $\Ht$
dependence for either $F_{QCD}$ or $k$, we simply multiply their
average values for $\Ht>200~\GeV$. Measured values of $k$ times
$F_{QCD}$ are given in Table~\ref{tab:Fnw}. The procedure by which
$F_{QCD}$ is determined as a function of $\Ht$ is important because
the ratios between the four different regions of the
 $\met$ and $I$ kinematic plane, calculated in separate ranges of $\Ht$
  and then averaged, does not necessarily correspond to the same ratios taken while integrating over
the full $\Ht$ range.
The uncertainties on the 1 and 2 jet events are conservatively taken as the
difference between the central value and the result of the straight product of $F_{QCD}$ and $k$.
The straight product corresponds to ignoring the $\Ht$ dependence as well as any other variable's
dependence of $F_{QCD}$ and $k$.

\begin{table}
\caption{\label{tab:Fnw}QCD fraction, given by
Equation~\ref{eq:FQCD}, and the $\Ht$-averaged product of $k$ and
the QCD fraction.}
\begin{ruledtabular}
\begin{tabular}{cccccc}
Jet multiplicity & 1 jet & 2 jet & 3 jets & $\geq$ 4
jets & $\geq 3$ jets \\ \hline \multicolumn{6}{c}{Muons}  \\
\hline
F$_{QCD}$ & $0.03\pm0.002$ & $0.039\pm0.004$ & $0.023\pm0.011$ & $0.146\pm0.088$ & $0.044\pm0.016$ \\
$\langle k_{\mu}\cdot$F$_{QCD}\rangle$ & $0.02\pm0.07$ & $0.05\pm0.51$ & $0.01\pm0.01$ & $0.09\pm0.07$ & $0.03\pm0.02$\\
\hline \hline
\multicolumn{6}{c}{Electrons}  \\ \hline
F$_{QCD}$ & $0.145\pm0.007$ & $0.177\pm0.010$ & $0.135\pm0.032$ & $0.163\pm0.051$ & $0.145\pm0.028$ \\
$\langle k_e\cdot$F$_{QCD}\rangle$ & $0.24\pm0.31$ & $0.41\pm0.12$ & $0.16\pm0.046$ & $0.20\pm0.07$ & $0.17\pm0.04$ \\
\end{tabular}
\end{ruledtabular}
\end{table}

\subsection{\label{sec:DY}Drell-Yan Background}
Drell-Yan events can enter the sample when they are produced with
jets and one muon is identified as the primary muon while the second
muon is close enough to a jet to be tagged. Residual Drell-Yan
background that is not removed by the dimuon and sequential decay
rejection described in Section~\ref{sec:Evsel}, is estimated from
the data. We use events inside the $Z$-mass window
($76-106~\GeVcc$), which are otherwise removed from the analysis by
the $Z$-mass cut, to measure the number of events that would pass
all our selection requirements including the SLT tag, $N^{\rm
tags}_{\rm inside}$. Because of the limited sample size of $Z$+jets
events, we use $Z+0~{\rm jet}$ events without the $\met$ and $\Ht$
requirements to find the ratio of events outside the $Z$-mass window
to those inside the window, $R^{{\rm out}/{\rm in}}_{Z/\gamma^*}$,
and a first-order estimate of the number of expected Drell-Yan
events outside of the $Z$-mass window is calculated as:

\begin{equation}
N_{DY} = N^{\rm tags}_{\rm inside} \cdot R^{{\rm out}/{\rm in}}_{Z/\gamma^*}.
\label{eq:NDY}
\end{equation}

\noindent This estimate assumes that $R^{{\rm out}/{\rm
in}}_{Z/\gamma^*}$ does not depend on the number of jets in the
event.  We assign a systematic uncertainty of 33\% for this
assumption based on the largest deviation between $R^{{\rm out}/{\rm
in}}_{Z/\gamma^*}$ for {\tt ALPGEN} $Z/\gamma^*$ plus zero jet
events compared with 1, 2 or $\ge$3 jets events.

The first-order estimate is then corrected by the relative
efficiency inside and outside the $Z$-mass window of the $\met$,
$\Ht$, and SLT-jet requirements, which we measure using
$Z/\gamma^*$+jets Monte Carlo events. The Drell-Yan background estimates are listed
in the sixth line of Table~\ref{tab:results}.

\subsection{\label{sec:MCBkg}Other Backgrounds}
Remaining background sources are due to $WW$, $WZ$, $ZZ$,
$Z\rightarrow\tau\tau$ and single top production.  Diboson events
can enter the sample when there are two leptons from a $Z$ and/or a
$W$ decay and jets.  One lepton passes the primary lepton
requirements while the second is available to pass the SLT
requirement if it is close to a jet.  The $\met$ in these events can
either come from a $W$-boson decay or from an undetected lepton in a
$Z$-boson decay. $Z\rightarrow\tau\tau$ events can enter the sample
when the $Z$ is produced in association with jets and one $\tau$
decays to a high-$\Pt$ isolated electron or muon, while the second
$\tau$ produces an SLT muon in its decay.  Electroweak single top
production gives rise to an event signature nearly identical to
$\ttbar$ when there are additional jets from gluon radiation.

\begin{table}[htbp]
\caption{\label{tab:Othbkg} Summary of the expected number of
background events for those sources derived from Monte Carlo
simulations, and the cross sections used in Equation~\ref{eq:MCBKG}.
The quoted uncertainties come from the respective Monte Carlo sample
sizes and the uncertainty on the theoretical cross sections.}
\sans
\begin{ruledtabular}
\begin{tabular}{cccccc}
 & 1 jet & 2 jets & 3 jets & $\ge$ 4 jets & $\ge$ 3 jets \\ \hline
 $WW$~\cite{Campbell}  & 0.64$\pm$0.15 & 0.99$\pm$0.18 & 0.12$\pm$0.07 &
 0.029$\pm$0.033 & 0.15$\pm$0.08\\
 $WZ$~\cite{Campbell}  & 0.11$\pm$0.07 & 0.22$\pm$0.09 & 0.03$\pm$0.04 &
 0.003$\pm$0.006 & 0.03$\pm$0.04 \\
 $ZZ$~\cite{Campbell}  & 0.013$\pm$0.010 & 0.025$\pm$0.015 & 0.007$\pm$0.007 &
 0.004$\pm$0.004 & 0.010$\pm$0.010\\
 $Z\ra \tptm$~\cite{CDFWZPRL} & 0.34$\pm$0.16 & 0.10$\pm$0.05 & 0.006$\pm$0.003 &
 0.002$\pm$0.001 & 0.008$\pm$0.004 \\
 Single top~\cite{Harris}  & 0.50 $\pm$ 0.03  &  0.94 $\pm$ 0.06 &
 0.15 $\pm$ 0.01 & 0.035 $\pm$ 0.003 & 0.19 $\pm$ 0.01\\
\end{tabular}
\end{ruledtabular}
\end{table}

None of the above background sources are completely accounted for by
the application of the tag matrix to the pretag event sample because
these backgrounds have a significant source of muons from, for
instance, $W$ and $Z$ decay. Therefore, we independently estimate
their contributions to the background using Monte Carlo samples
normalized to the cross sections referenced in
Table~\ref{tab:Othbkg}. In modeling the SLT tagging of such events
in the Monte Carlo samples, we explicitly exclude the mistag
contribution which is taken in to account in the application of the
tag matrix to the pretag sample. The background for each source is
estimated as:

\begin{equation}
N_i=\sigma_i\cdot A_i \cdot \epsilon_{{\rm tag},i} \cdot \int
\cal{L}{\rm dt},
\label{eq:MCBKG}
\end{equation}

\noindent where $\sigma_i$ is the theoretical cross section for the
particular background source, $A_i$ is the acceptance for passing
the pretag event selection, $\epsilon_{{\rm tag},i}$ is the SLT
tagging efficiency and $\int\cal{L}{\rm dt}$ is the integrated
luminosity of the overall data sample. The expected background
contributions are shown, as a function of jet multiplicity, in
Table~\ref{tab:Othbkg}.

\section{\label{sec:Acc}Total {\lowercase{\boldmath $t \bar t$}} Acceptance}
We factorize the efficiency for identifying $\ttbar$ events into the
geometric times kinematic acceptance and the SLT tagging efficiency.
The acceptance includes all the cuts described in
Sections~\ref{sec:Wplusjets} as well as the invariant mass cut
described in Section~\ref{sec:Evsel}, and is evaluated assuming a
top mass of 175 GeV/c$^2$. The tagging efficiency is the efficiency
for SLT-tagging at least one jet in events that pass the geometric
and kinematic selection. We describe each piece below.

\subsection{\label{subsec:geokin} Geometric and Kinematic Acceptance}
The acceptance is measured in a combination of data and Monte Carlo
simulations.  Simulations are done using the {\tt PYTHIA} Monte
Carlo program~\cite{Pythia}. The primary lepton identification
efficiency is measured in $Z$-boson decays acquired with a trigger
that requires a single high-$\Pt$ electron or muon.  The efficiency
is measured using the lepton from the $Z$-boson decay that is
unbiased by the trigger, and the identification efficiency in the
Monte Carlo sample is scaled to that measured in the
data~\cite{kinPRD}. The acceptance, as a function of the number of
identified jets above $15~\GeV$, is shown in Table~\ref{tab:Acc}.
These numbers include the measured efficiencies of the high-$\Pt$
lepton triggers.

\begin{table}[htbp]
\caption{\label{tab:Acc} Acceptance for $\ttbar$ events as a
function of jet multiplicity from {\tt PYTHIA} Monte Carlo sample,
corrected for the data/MC ratio for tight lepton ID efficiencies and
the primary lepton trigger efficiency. The uncertainties listed are
statistical only.}
 \sans
\begin{ruledtabular}
\begin{tabular}{cccccc}
 & $W$+ 1 jet & $W$ + 2 jets & $W$ + 3 jets & $W$ + $\ge$ 4 jets & $W$ + $\ge$ 3 jets \\ \hline
$W\rightarrow e\nu$ (\%)& 0.204$\pm$0.005 & 1.05$\pm$0.01 & 1.79$\pm$0.02 & 2.27$\pm$0.02 & 4.06$\pm$0.02\\
$W\rightarrow \mu\nu$ (CMUP) (\%)  & 0.095$\pm$0.003 & 0.501$\pm$0.007 & 0.861$\pm$0.007 & 1.12$\pm$0.01 & 1.98$\pm$0.01 \\
$W\rightarrow \mu\nu$ (CMX) (\%) & 0.045$\pm$0.002 & 0.235$\pm$0.006 & 0.388$\pm$0.007 & 0.507$\pm$0.008 & 0.90$\pm$0.01 \\
\end{tabular}
\end{ruledtabular}
\end{table}

\subsection{\label{subsec:slteff} SLT Efficiency}
\begin{figure}[htbp]
\begin{center}
\includegraphics[width=7.5cm]{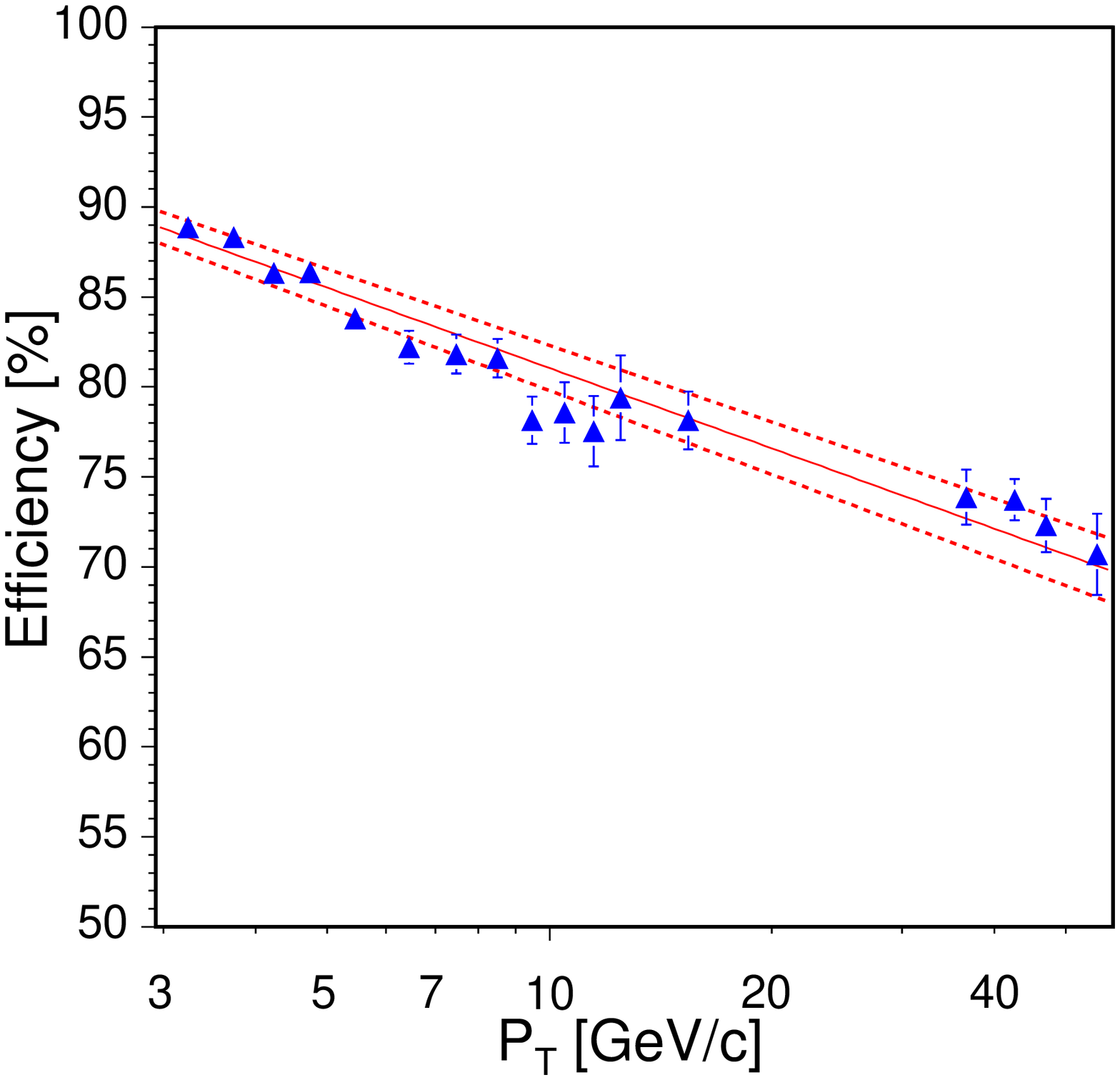}
\includegraphics[width=7.5cm]{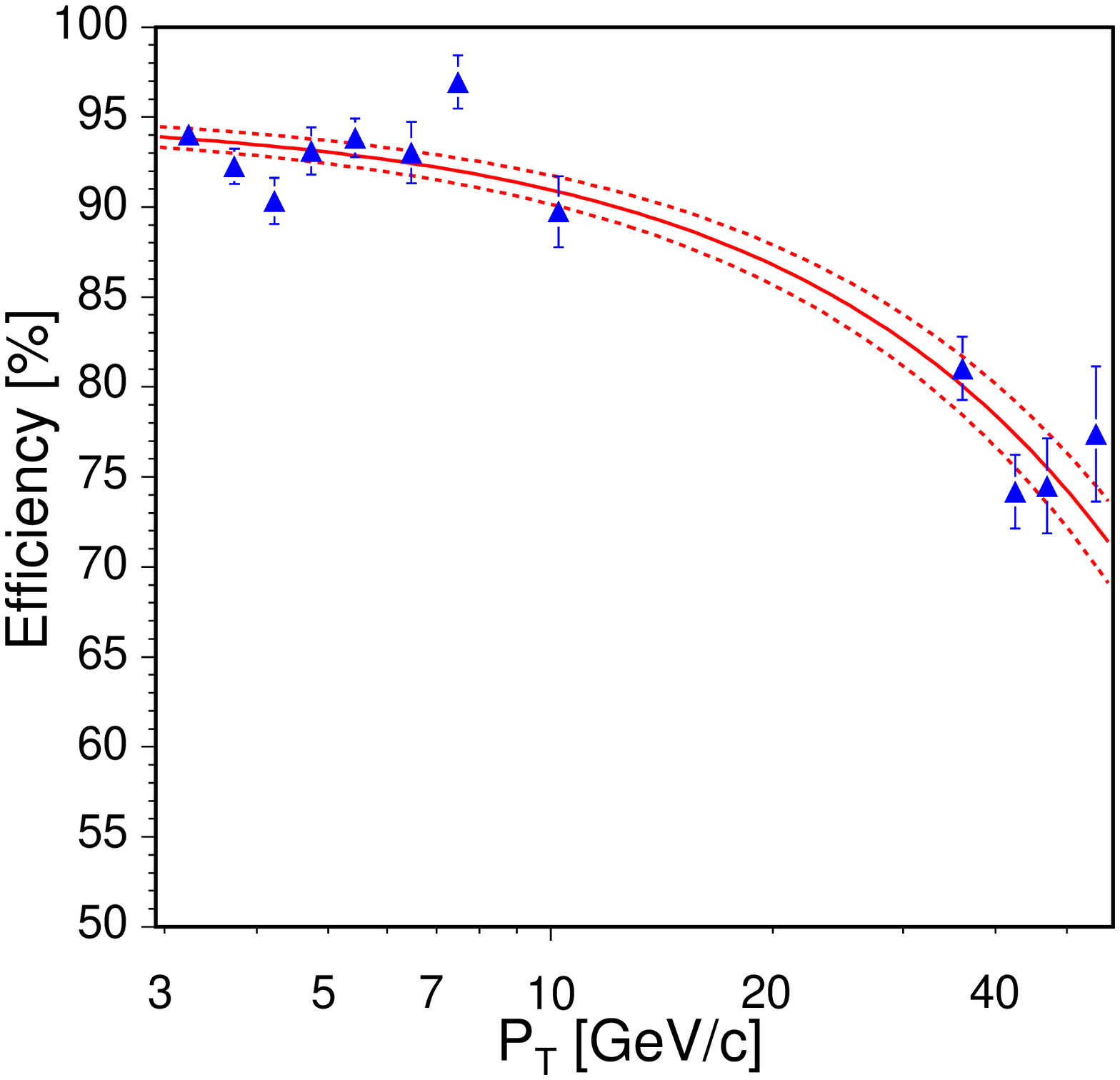}
\caption{The SLT efficiency for $|\eta|<0.6$ (left) and
$|\eta|\ge 0.6$ (right) as a function of $P_T$,
measured from $J/\psi$ and $Z$ data for $|L|<3.5$. The dotted lines
are the $\pm 1\sigma$ statistical uncertainty on the fit which is used in the
evaluation of the systematic uncertainty.}
\label{fig:eff}
\end{center}
\end{figure}

The efficiency for the reconstruction of the COT track is taken
directly from Monte Carlo simulation. The reconstruction
efficiencies of muon chamber track segments are also taken from the
simulation and scaled to the values measured in the data using the
lepton in $Z$-boson decays unbiased by the trigger. The muon
identification efficiency of the SLT algorithm is measured in data
using $J/\psi$ and $Z$ decays. We use events acquired with triggers
that demand a single muon, and use only the muon not biased by the
trigger. The efficiency is defined as the ratio of muons that
satisfy the SLT tagging requirement over the number of taggable
tracks attached to track segments in the muon chambers. In the
calculation of the efficiency a background linear in invariant mass
is subtracted from the $J/\psi$ and $Z$ peaks. The measured
efficiency vs.\@ $\Pt$ is shown in Figure~\ref{fig:eff} for muons
with $|\eta|<0.6$ (CMU and/or CMP) and for muons with
$0.6\le|\eta|\le 1.0$ (CMX). The decrease in efficiency with
increasing $\Pt$ is a result of non-Gaussian tails in the components
of $L$.

Since the efficiency measurement is dominated by isolated muons,
whereas the muons in $b$-jets tend to be surrounded by other tracks,
we have studied the dependence of the efficiency on the number of
tracks, $N_{\rm trk}$, above $1~\GeVc$ in a cone of $\Delta R=0.4$
around the muon track. We find no significant efficiency loss,
although the precision of the measurement is poor near $N_{\rm
trk}=6$, the mean expected in $\ttbar$ events.  We include a
systematic uncertainty to account for this by fitting the efficiency
vs.\@ $N_{\rm trk}$ to a linear function and evaluating this
function at the mean $N_{\rm trk}$ expected in $\ttbar$ events.  The
systematic uncertainty on the efficiency for at least one SLT tag in
a $\ttbar$ event from this effect is +0\%, $-$8\%.

The detector simulation does not properly reproduce the non-Gaussian
tails of the muon matching distributions.  Therefore the measured
efficiencies, shown in Figure~\ref{fig:eff}, are applied directly to
a generated muon in the Monte Carlo sample when evaluating the
efficiency for tagging a $\ttbar$ event. This accounts for tagging
of semileptonic heavy flavor decays in $\ttbar$ events (including
charm decays from $W\!\rightarrow\!c\bar{s}$). Events from $\ttbar$
can also be mistagged when a tag results from a fake muon or a
decay-in-flight. We account for this effect in the tagging
efficiency evaluation by allowing events that are not tagged by
muons from heavy flavor decays to be tagged by other charged tracks
using the tagging probabilities from the tag matrix, as described in
Section~\ref{sec:Bak}. Since the heavy flavor component of the
tagging efficiency has already been accounted for, the generic-track
tagging probabilities are corrected downwards for the measured 21\%
heavy flavor component of the tag matrix (Section~\ref{sec:Whf}).
The overall efficiency for finding one or more SLT tags in a
$t\bar{t}$ event (``tagging efficiency") is shown in
Table~\ref{tab:effresults}. Mistags account for approximately 25\%
of the $\ttbar$ tagging efficiency. Because a small portion of the
integrated luminosity was accumulated before the CMX was fully
functional, we break the efficiency into pieces with and without the
CMX.  This is taken into account in the final cross section
denominator. The total $t\bar{t}$ detection efficiency is the
product of the
 acceptance and the tagging efficiency.

As noted above, the SLT efficiency has been parameterized using
muons that tend to be isolated from other activity.  To further
check that this efficiency measurement is representative of muons in
or near jets, we use a high-purity $b\bar{b}$ sample, derived from
events triggered on $8~\GeV$ electrons or muons. These events are
enriched in semileptonic $b$-hadron decays. To select this sample,
we require that the events have two jets above $15~\GeV$.  One jet
must be within $\Delta R=0.4$ of the primary electron or muon (the
``lepton jet"). For jets associated with muons, the energy is
corrected to account for the muon $\Pt$. The second jet (the ``away
jet") in the event is chosen as the jet above $15~\GeV$ with maximum
separation in azimuth ($\ge$2 radians) from the lepton jet. Both
jets are required to have a secondary vertex reconstructed and
tagged by the SecVtx algorithm~\cite{SVX} (``SecVtx-tagged"). This
results in a $b\bar{b}$ sample with a purity of approximately
95\%~\cite{Lannon}. We measure the SLT acceptance times efficiency
for semileptonic decays to muons in the away jet in a {\tt HERWIG}
dijet Monte Carlo sample. Monte Carlo events are subject to the same
event selection, as described above, used for the $b\bar{b}$ data
sample. The efficiency parametrization measured from the data is
applied in the same way as in the $\ttbar$ Monte Carlo sample. The
derived efficiency times acceptance per $b$-jet is applied to the
data to predict the number of SLT tags in the away jet.  There are
7726 SecVtx-tagged away jets in which the lepton jet is from a muon
and 2233 in which it is from an electron. In these events we predict
$388\pm54$ tags in the away jet opposite a muon jet and $116\pm17$
tags in the away-jet opposite an electron jet. We find 353 and 106
respectively.  We conclude that the efficiency for SLT-tagging muons
from semileptonic decays of heavy flavor in jets is well modeled by
our simulation.

\begin{table}[htbp]
\caption{\label{tab:effresults} $\ttbar$ event tagging efficiency
for SLT muons as a function of jet multiplicity from {\tt PYTHIA}
Monte Carlo sample. Uncertainties are statistical only.}
\sans
\begin{ruledtabular}
\renewcommand{\arraystretch}{1.25}
\begin{tabular}{cccccc}
 & $W$ + 1 jet & $W$ + 2 jets & $W$ + 3 jets & $W$ + $\ge$ 4 jets & $W$ + $\ge$ 3jets\\ \hline
$W\rightarrow e\nu$ w/CMX (\%) & $9.5\pm0.8$ & $13.1\pm0.4$ & $14.7\pm0.3$ & $15.9\pm0.3 $ & $15.4\pm0.2$ \\
$W\rightarrow e\nu$ w/o CMX (\%)& $6.7\pm0.7$ &$10.3\pm0.4$ & $11.5\pm0.3$ & $12.4\pm0.3$ & $12.0\pm0.2$ \\
\hline\hline
$W\rightarrow\mu\nu$ w/CMX (\%) & $7.2\pm0.8$ & $12.3\pm0.5$ & $13.3\pm0.3$ & $16.1\pm0.3$ & $14.9\pm0.2$ \\
$W\rightarrow\mu\nu$ w/o CMX (\%) & $5.0\pm0.8$ & $9.6\pm0.5$ & $10.3\pm0.4$ & $12.8\pm0.3$ & $11.7\pm0.3$ \\
\end{tabular}
\renewcommand{\arraystretch}{1.00}
\end{ruledtabular}
\end{table}

\section{\label{sec:sys}Systematic Uncertainties}
Systematic uncertainties in this analysis come from uncertainties in
the Monte Carlo modeling of the acceptance, knowledge of the SLT
tagging efficiency, the effect on the acceptance of the uncertainty
on the jet energy calibration, uncertainties on the background
predictions, and the uncertainty on the luminosity.

Uncertainties in the Monte Carlo modeling of acceptance include
effects of parton distribution functions (PDFs), initial-state
radiation (ISR), final-state radiation (FSR), and the calibration of
the measured jet energy. These are estimated by comparing different
choices for PDFs, varying ISR, FSR and the jet energy in the Monte
Carlo programs and comparing the results from the {\tt PYTHIA}
generator with those from {\tt HERWIG}.  A complete description of
the evaluation of these uncertainties appears in~\cite{kinPRD}. The
total systematic uncertainty on the acceptance due to these factors
is $\pm$6.1\%. Possible variations of the lepton ID efficiency in
events with multiple jets are an additional source of systematic
uncertainty on the acceptance.  We use a data to Monte Carlo scale
factor for the lepton ID efficiency that is taken from $Z\rightarrow
ee$ and $Z\rightarrow\mu\mu$ data and Monte Carlo samples. These
samples contain predominantly events with no jets. A 5\% systematic
uncertainty on the scale factor is estimated by convoluting the
scale factor itself, measured as a function of $\Delta R$ between
the lepton and the nearest jet, with the $\Delta R$ distribution of
leptons in $\ge$3 jet $t\bar{t}$ events~\cite{kinPRD}. Adding the
uncertainties in quadrature gives a total Monte Carlo modeling
systematic uncertainty on the acceptance of $\pm$8.0\% .

There are several factors that contribute to the systematic
uncertainty on the SLT tagging efficiency. The uncertainty due to
our knowledge of the $\Pt$ dependence
 is determined by varying the efficiency curves used
in the $t\bar{t}$ Monte Carlo sample according to the upper and
lower bands in Figures~\ref{fig:eff}.  We find that the tagging
efficiency for $t\bar{t}$ changes by $\pm$1\% from its central
value.  An additional source of systematic uncertainty for the
tagging efficiency comes from the fact that we implicitly use the
Monte Carlo tracking efficiency for taggable tracks.  As these
tracks can be in dense environments in or near jets, we expect the
efficiency to be somewhat less than for isolated tracks. Studies
done by embedding Monte Carlo tracks in jets in both data and Monte
Carlo events indicate that the Monte Carlo tracking efficiency in
dense environments is a few percent higher than in data. We assign a
$\pm$5\% systematic uncertainty to the tagging efficiency for this
effect. As described in Section~\ref{subsec:slteff} the systematic
uncertainty due to the modeling of the isolation dependence of the
tagging efficiency is $+$0\%, $-$8\%. Finally, the statistical
uncertainty on the measurement of the SLT tagging efficiency in
$\ttbar$ events, differences between {\tt PYTHIA} and {\tt HERWIG},
the uncertainty on the semileptonic branching fraction for B mesons
and the estimation of the heavy flavor content of the mistag matrix
also contribute to the systematic uncertainties. Adding these
contributions in quadrature gives an overall systematic uncertainty
for the tagging efficiency of $+$8\%, $-$11\%. Note that the
uncertainty on the tagging efficiency affects also the backgrounds
determination. The reason is that $\ttbar$ events need to be
subtracted from the pretag sample which is used in
Equation~\ref{eq:Fake} to determine the $W$+jets background. We take
this effect into account when calculating the uncertainty on the
cross section.

Uncertainties on the tag matrix are determined by the level of
agreement between observed tags and predictions in a variety of
samples, as described in Section~\ref{sec:Bak}.  The uncertainty on
the $W$+fakes and $Wb\bar{b}+Wc\bar{c} + Wc$ prediction is
$\pm$10\%.

To determine the uncertainties on the QCD background prediction in
events with three or more jets, we define a control sample from the
$\met$ vs.\@ lepton isolation plane, where the primary lepton
isolation parameter $I$ is between 0.1 and 0.2 and the event has
$\met>20~\GeV$. After subtracting expected contributions from $W$
and $\ttbar$ events, all events in this region are expected to be
QCD.  We determine the systematic uncertainty on the QCD background
using the ratio of the observed over predicted number of events in
this control region, which should be 1.0. In the sample where the
primary lepton is a muon, we measure 0.5$\pm$0.4.  In the sample
where the primary lepton is an electron, we measure 0.8$\pm$0.2. A
50\% systematic uncertainty is assigned to the F$_{QCD}$ measurement
for muons and 20\% for electrons. We combine this with the
statistical uncertainty on F$_{QCD}$, the uncertainty on the
correction factor $k$, both given in Table~\ref{tab:Fnw} and the
10\% systematic uncertainty due to the application of the tag
matrix. The total QCD background uncertainty is $\pm$67\% and
$\pm$19\% for muons and electrons, respectively. These values are
determined taking into account the correlation between the estimate
of the QCD background and the estimate of the $W$+fakes and
$W$+heavy flavor backgrounds (Equations~\ref{eq:bkg0}
and~\ref{eq:bkg1}).  We add in quadrature the separate effects on
the cross section of the QCD uncertainties for electrons and muons.

The systematic uncertainty on the small Drell-Yan background is
dominated by its statistical uncertainty. We also include a 33\%
relative uncertainty to account for changes in the shape of the
Drell-Yan spectrum with the number of jets in the event, as
described in Section~\ref{sec:DY}. Uncertainties on the Monte Carlo
background predictions come from uncertainties in the cross sections
for the various processes and from the event sizes of the Monte
Carlo samples.

The systematic uncertainties and the corresponding shift of the measured
cross section value are summarized in Table~\ref{tab:sys}.

\begin{table}[htbp]
\caption{\label{tab:sys} Summary of systematic uncertainties.  The
shift $\Delta\sigma_{\ttbar}$ of the measured cross section value
assumes the cross section calculated in Section~\ref{sec:results}.}
\begin{ruledtabular}
\begin{tabular}{lcc}
Source & Fractional Sys.\@ Uncert.\@ (\%) & $\Delta\sigma_{\ttbar}$
(pb)\\ \hline
Acceptance Modeling & $\pm$8 & \multirow{2}{*}{  {\Large \{ $^{+1.10}_{-0.70}$} } \\
SLT Tagging Efficiency & ${+8},{-11}$ & \\
Tag Matrix Prediction & $\pm$10  & $\pm$0.68 \\
QCD Prediction & $\pm$19 (e) $\pm$67 ($\mu$) & $\pm$0.14\\
Drell-Yan and other MC backgrounds & $\pm$19 & $\pm$0.05\\
Luminosity &  $\pm$6 & $\pm$0.32\\ \hline\hline
\multirow{2}{*}{Total Systematic Uncertainty} & & \multirow{2}{*}{\Large{$^{+1.3}_{-1.0}$}} \\
 & & \\
\end{tabular}
\end{ruledtabular}
\end{table}

\section{\label{sec:results}Results}
Table~\ref{tab:results} shows a summary of the background estimates
and the number of SLT tagged events as a function of the number of
jets.  A ``tagged event" is an event with at least one tagged jet.
The total background and the $t\bar{t}$ expectation are also listed.
The line labeled ``Corrected Background'' corresponds to the
background after correcting for the $\ttbar$ content of the pretag
sample, as described in Section~\ref{sec:Whf}.

We calculate the cross section as:

\begin{equation}
\sigma_{t\bar{t}}=\frac{N_{\rm obs}-N_{\rm bgnd}}{A_{t\bar{t}}\cdot
\int\mathcal{L} dt},
\label{eq:xsec}
\end{equation}

\noindent where $N_{\rm obs}$ is the number of events with $\ge 3$ jets
that are tagged with at least 1 SLT, $N_{\rm bgnd}$ is the corrected
background and $A_{t\bar{t}}$ is the total acceptance (geometrical
acceptance times kinematic acceptance times tagging efficiency),
taken from Tables~\ref{tab:Acc} and ~\ref{tab:effresults}.  For
events with three or more jets, the total denominator is 1.98 $\pm$ 0.28
pb$^{-1}$.

From the number of candidate events with three or more jets, we find
a total $\ttbar$ production cross section of

\begin{center}
$\sigma(\ppbar\rightarrow\ttbar)=5.3\pm3.3~^{+1.3}_{-1.0} ~\rm pb, $
\end{center}

\noindent where the first uncertainty is statistical and the second
is systematic.  This cross section value uses acceptances and
tagging efficiencies appropriate for a top mass of $175~\GeVcc$.
The acceptances and efficiencies, and therefore the calculated cross
section, change slightly for other assumed top masses.  The
calculated cross section is 1\% higher assuming a top mass of $170~\GeVcc$,
and 5\% lower assuming a top mass of $180~\GeVcc$.

\begin{figure}[htbp]
\begin{center}
\includegraphics[width=4.0in]{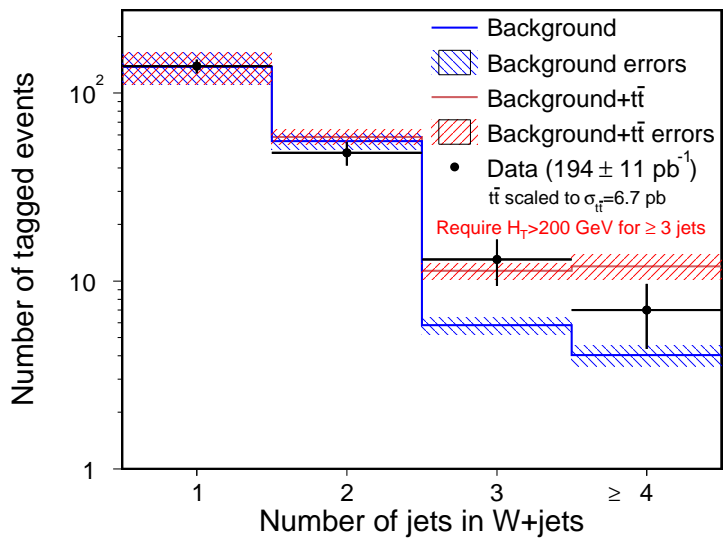}
\caption{The expected background and observed tags in $W+$ 1, 2, 3
and 4 or more jet events.  The background is corrected for the
$\ttbar$ content of the pretag sample.} \label{fig:njets}
\end{center}
\end{figure}

Figure~\ref{fig:njets} shows the number of tags in $W+1,2,3, \ge 4$
jet events together with the histograms representing the total
corrected background with and without the $t\bar t$ signal
expectation, based on the theoretical cross section of $6.7~$pb for
$M_{\rm top}=175~\GeVcc$.

\begin{figure}[htbp]
\begin{center}
\includegraphics[width=4.0in]{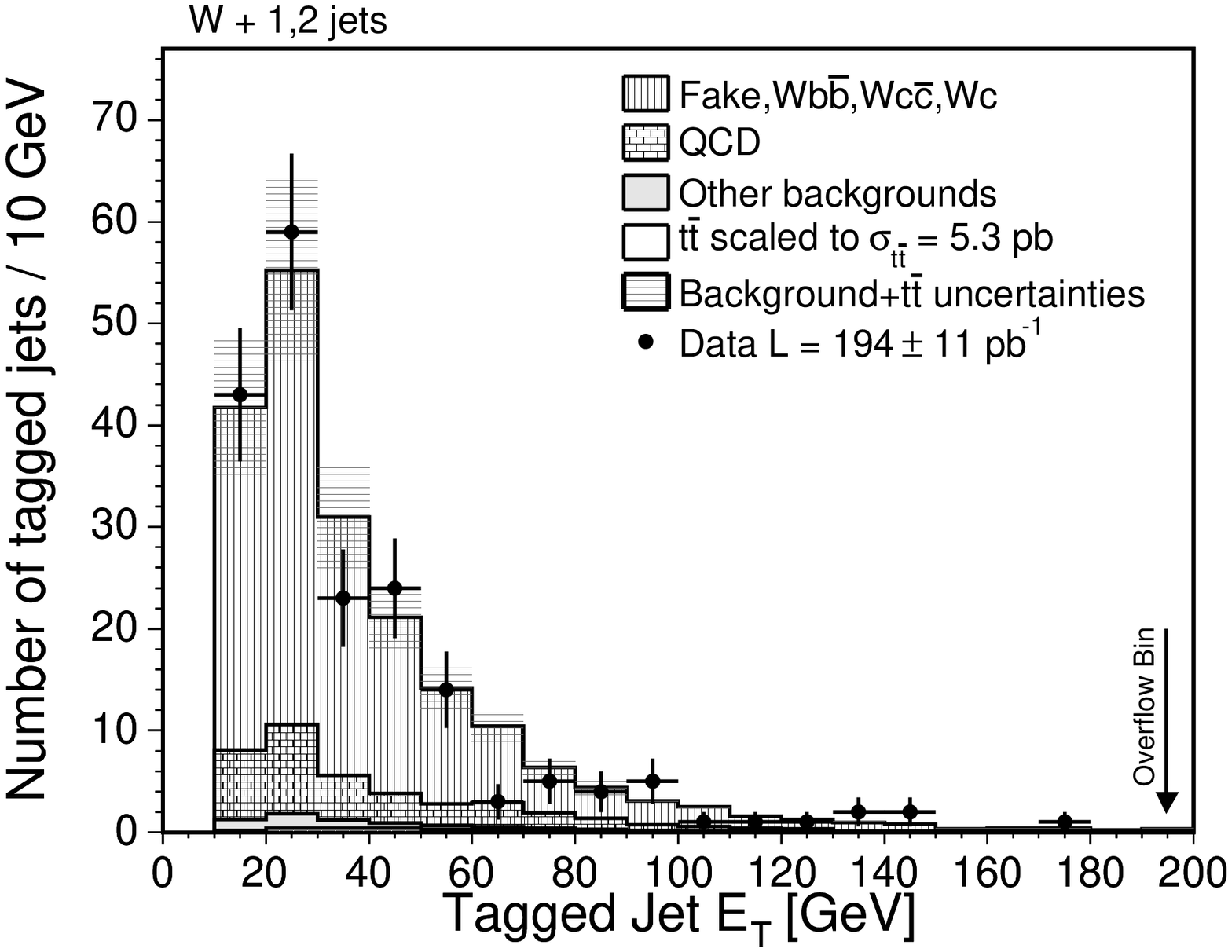}
\includegraphics[width=4.0in]{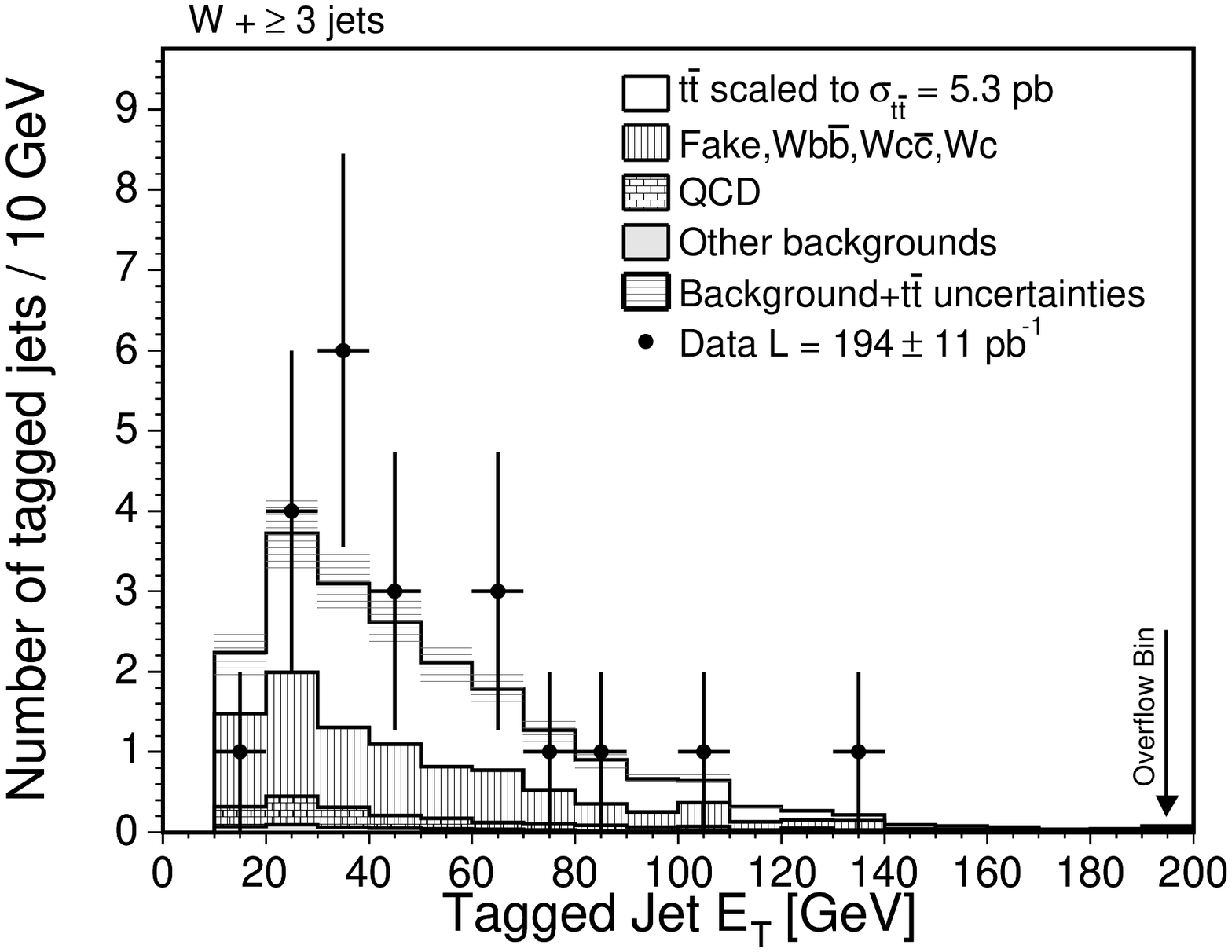}
\caption{Comparison of the jet $\Et$ distributions for tagged events
and for expectations from fakes, QCD and $\ttbar$ events. The
upper plot is for $W$+1 and 2 jet events and the lower plot for
$W+\ge 3$ jet events.} \label{fig:jetet}
\end{center}
\end{figure}

We examine a number of kinematic distributions of the tagged events
and compare with expectations based on the measured signal plus
background.  Figure~\ref{fig:jetet} shows the $\Et$ distribution of
the tagged jets in $W$+1 and 2 jets (combined), and in the signal
region of $W$ plus three or more jets.  The $W$+1 and 2 jet
data-Monte Carlo comparison has a Kolmogorov-Smirnov test (KS)
probability of 41\%, and the three or more jet comparison has a KS
probability of 82\%.

\begin{figure}[htbp]
\begin{center}
\includegraphics[width=4.0in]{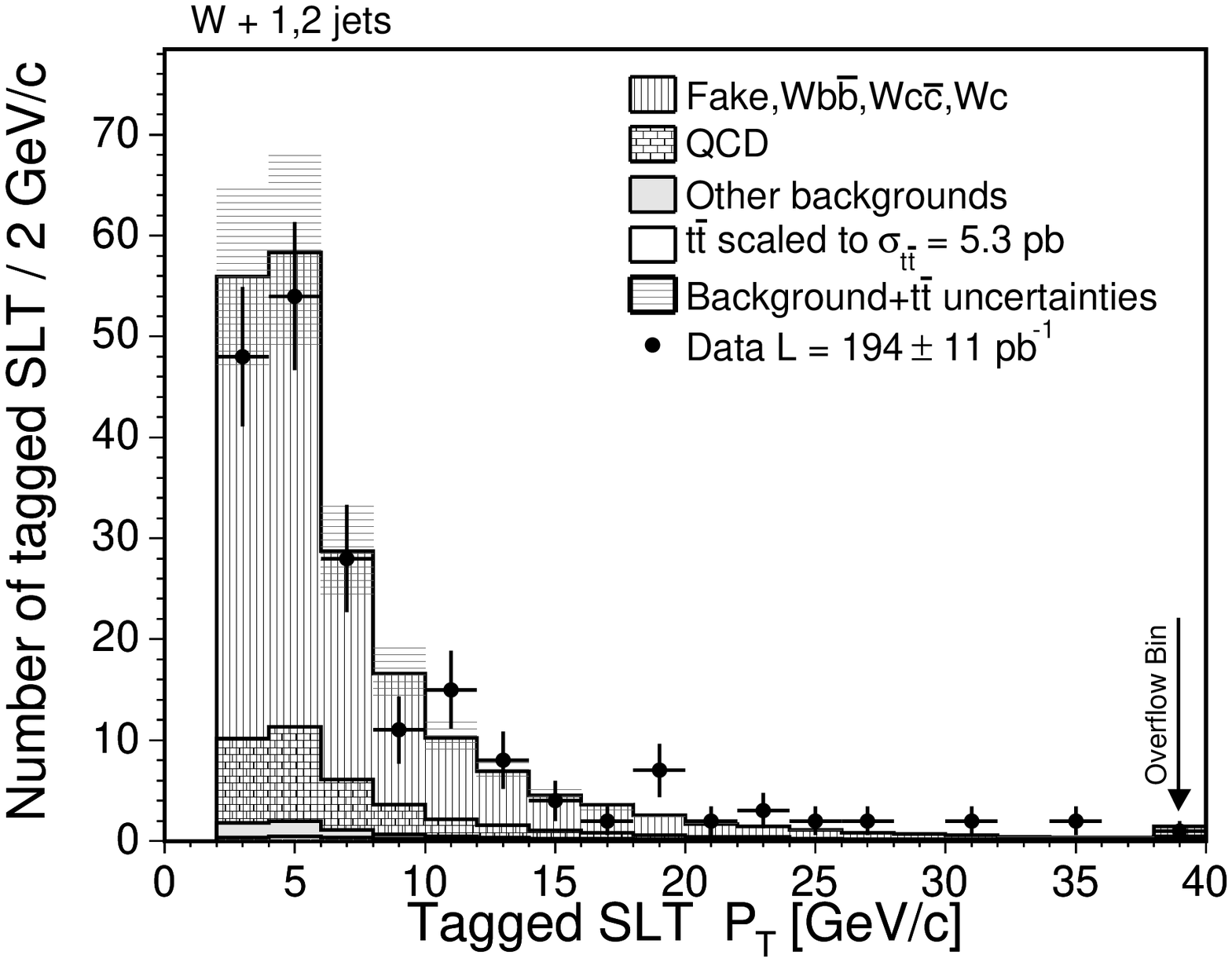}
\includegraphics[width=4.0in]{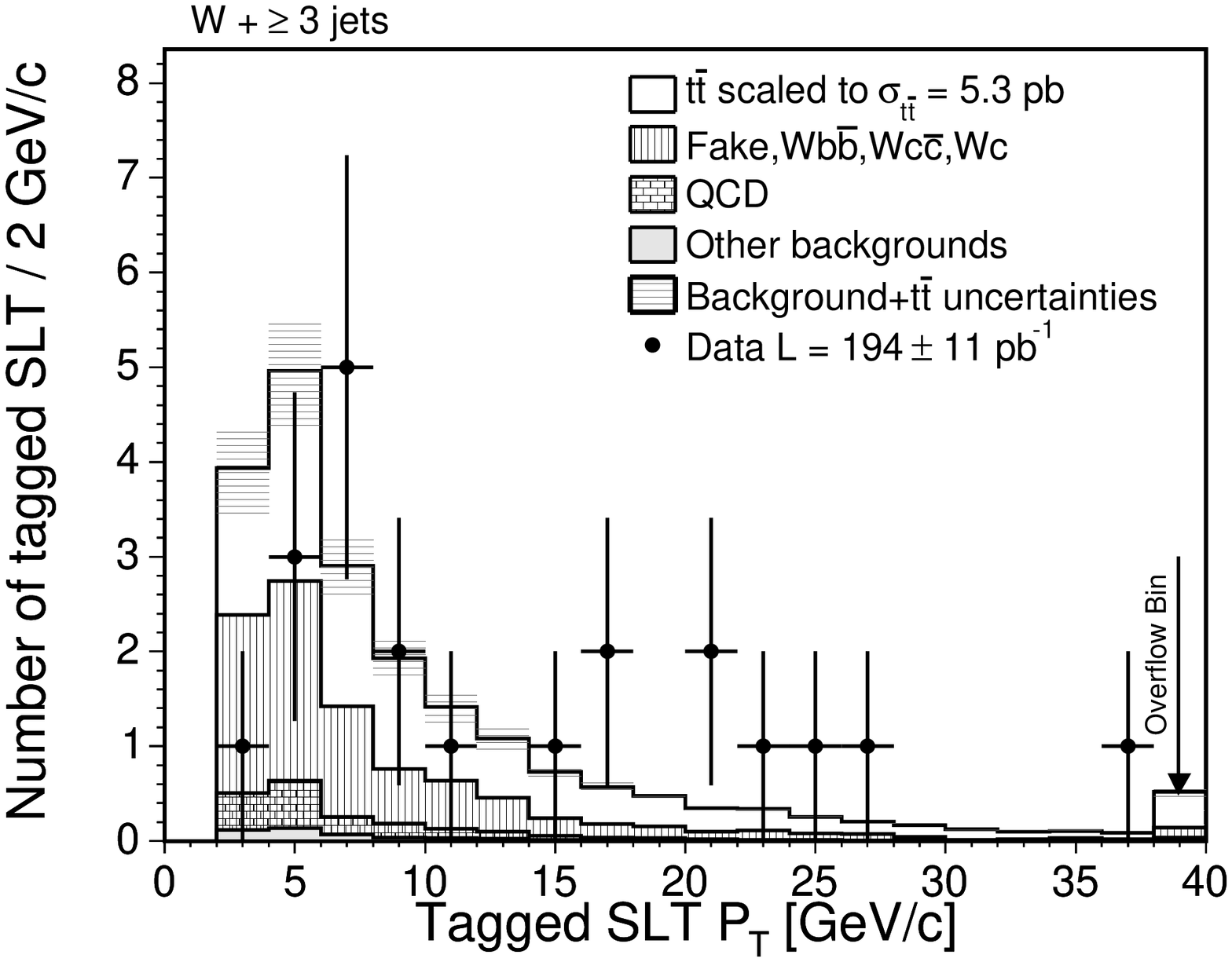}
\caption{$P_T$ of the SLT tags.  The upper plot is for $W$+1 and 2
jet events and the lower plot for $W+\ge 3$ jet events.}
\label{fig:tagPt}
\end{center}
\end{figure}

Figure~\ref{fig:tagPt} compares the $\Pt$ distribution of muons
identified as SLT tags with expectations from $\ttbar$ plus
backgrounds.  The KS probabilities are 6\% for $W$+1 and 2 jet
comparison and 5\% for the three or more jet comparison.

\begin{figure}[htbp]
\begin{center}
\includegraphics[width=4.0in]{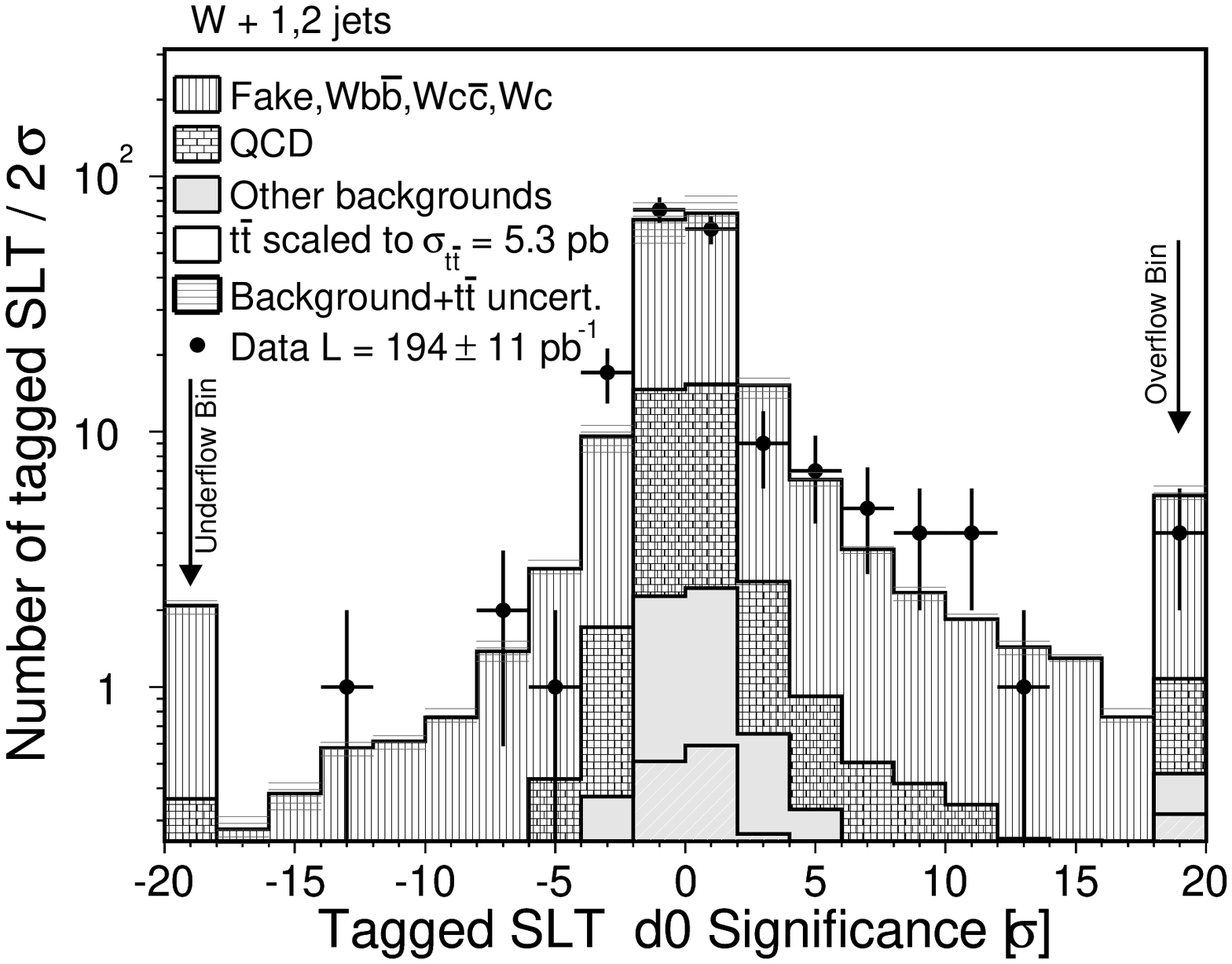}
\includegraphics[width=4.0in]{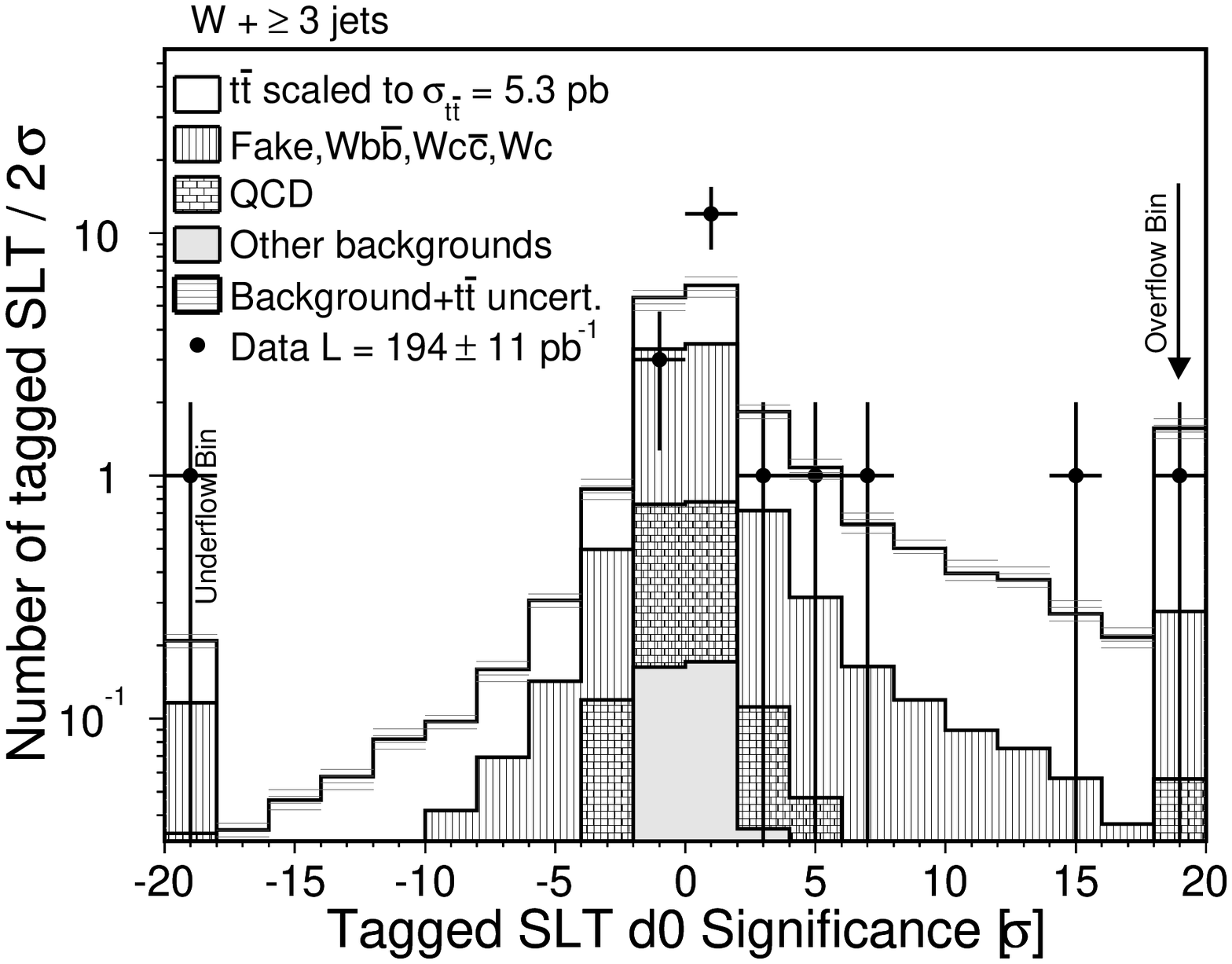}
\caption{The impact parameter (d$0$) significance for tagged events,
compared with expectations from backgrounds plus $\ttbar$. The upper
plot is for $W$+1 and 2 jet events and the lower plot for $W+\ge 3$
jet events.} \label{fig:d0sig}
\end{center}
\end{figure}

Finally, Figure~\ref{fig:d0sig} shows the impact parameter
significance, defined as the impact parameter divided by its
uncertainty, for the SLT tracks and the expectation from signal plus
background.  The sign of the impact parameter is defined according
to whether the track trajectory crosses the jet axis in front of or
behind the event primary vertex.  The long-lived component from
semi-leptonic $b$-hadron decays is readily apparent in the shape of
the positive impact parameter distribution. The KS probabilities are
12\% for $W$+1 and 2 jet comparison and 23\% for the three or more
jet comparison. (Note that Figures~\ref{fig:jetet}, \ref{fig:tagPt}
and \ref{fig:d0sig} contain 21 entries since one of the events has
two jets tagged with SLT.)

\section{\label{sec:concl}Conclusions}
We have measured the total cross section for $\ttbar$ production
through the decay of top pairs into an electron or muon plus
multiple jets.  We separate signal from background by identifying
semileptonic decays of $b$ hadrons into muons. The measured $\ttbar$
production cross section is $5.3\pm3.3$$^{+1.3}_{-1.0}$ pb,
consistent with the expectation of 6.7~pb for standard model
production and decay of top quark pairs with a mass of $175~\GeVcc$.
Distributions of Jet $\Et$ and impact parameter significance for the
tagged events, and the distributions of the $\Pt$ of the tags, are
also consistent with standard model expectations.

The sensitivity of this analysis to test non-standard model $\ttbar$
production or decay mechanisms is limited by the statistical
uncertainty.  The combination of this measurement with other CDF~II
measurements~\cite{combXsec} will yield a significantly more precise
value. Future measurements with the full Run~II dataset of
$4-8$~fb$^{-1}$ will provide further factors of approximately four
to six in statistical precision.  At the same time, significantly
larger datasets will provide avenues for reduction of the systematic
uncertainties through such things as better understanding of the tag
rate in $W$+jets events and direct measurement of the tagging
efficiency for semileptonic $b$-hadron decays.

\section{Acknowledgments}
We are grateful to Tim Stelzer for help with studies of the heavy
flavor content of $\gamma$+jets and $W$+jets events using {\tt
MADEVENT}. We thank the Fermilab staff and the technical staffs of
the participating institutions for their vital contributions. This
work was supported by the U.S. Department of Energy and National
Science Foundation; the Italian Istituto Nazionale di Fisica
Nucleare; the Ministry of Education, Culture, Sports, Science and
Technology of Japan; the Natural Sciences and Engineering Research
Council of Canada; the National Science Council of the Republic of
China; the Swiss National Science Foundation; the A.P. Sloan
Foundation; the Bundesministerium f\"ur Bildung und Forschung,
Germany; the Korean Science and Engineering Foundation and the
Korean Research Foundation; the Particle Physics and Astronomy
Research Council and the Royal Society, UK; the Russian Foundation
for Basic Research; the Comision Interministerial de Ciencia y
Tecnologia, Spain; and in part by the European Community's Human
Potential Programme under contract HPRN-CT-2002-00292, Probe for New
Physics.


\begin{thebibliography}{99}
\bibitem{MET} We use a ($z,\phi, \theta$) coordinate system where the $z$-axis
    is in the direction of the proton beam, and $\phi$ and $\theta$ are the azimuthal
    and polar angles respectively.  The pseudo-rapidity, $\eta$, is defined as
    $-\ln(\tan\frac{\theta}{2})$.  The transverse momentum of a charged particle is
    $\Pt = P\sin\theta$, where $P$ represents the measured momentum of the charged-particle
    track.  The analogous quantity using calorimeter energies, defined as $E_T =
    E\sin\theta$,
    is called transverse energy. The missing transverse energy is defined as
    $\met = -\mid\sum_i E_T^i\hat{n}_i\mid$ where $E_T^i$ is the
    magnitude of the transverse energy contained in each calorimeter
    tower $i$ in the pseudo-rapidity region $|\eta|<3.6$ and
    $\hat{n}_i$ is the direction unit vector of the tower in the
    plane transverse to the beam direction.
\end{thebibliography}
\end{document}